\documentclass[12pt,draftclsnofoot,journal,onecolumn]{IEEEtran}
\usepackage{graphicx}
\usepackage{amsmath}

\usepackage{amsfonts}
\usepackage{amssymb}
\usepackage{array}
\newcommand{\PreserveBackslash}[1]{\let\temp=\\#1\let\\=\temp}
\newcolumntype{C}[1]{>{\PreserveBackslash\centering}p{#1}}
\newcolumntype{R}[1]{>{\PreserveBackslash\raggedleft}p{#1}}
\newcolumntype{L}[1]{>{\PreserveBackslash\raggedright}p{#1}}
\usepackage[usenames]{color}
\usepackage{colortbl,booktabs}
\usepackage{tabularx}
\usepackage{multicol}
\usepackage{booktabs}
\usepackage[Symbol]{upgreek}
\usepackage{subfigure}
\usepackage{bm}
\usepackage{cite}
\usepackage{balance}
\usepackage[ruled,vlined]{algorithm2e}

\usepackage{threeparttable}
\usepackage{amsmath}
\usepackage{dcolumn}
\usepackage{multirow}
\usepackage{amsfonts}

\newcolumntype{d}[1]{D{.}{.}{#1}}

\begin{document}
	
	\bibliographystyle{IEEEtran} 
	\title{Hierarchical Beam Training for Extremely Large-Scale MIMO: From Far-Field to Near-Field 
	}
	
	\author{Yu Lu,~\IEEEmembership{Student Member,~IEEE}, Zijian Zhang,~\IEEEmembership{Student Member,~IEEE}, and Linglong Dai,~\IEEEmembership{Fellow,~IEEE}

		\vspace{-2em}
		\thanks{All authors are with the Department of Electronic Engineering, Tsinghua University as well as	Beijing National Research Center for Information Science and Technology (BNRist), Beijing 100084, China (e-mails: y-lu19@mails.tsinghua.edu.cn, zhangzj20@tsinghua.edu.cn, daill@tsinghua.edu.cn).}
	}
	
	\maketitle
	\vspace{-0mm}
	\begin{abstract}
Extremely large-scale MIMO (XL-MIMO) is a promising technique for future 6G communications. The sharp increase in the number of antennas causes electromagnetic propagation to change from far-field to near-field. Due to the near-field effect, the exhaustive near-field beam training at all angles and distances requires very high overhead. The improved fast near-field beam training scheme based on time-delay structure can reduce the overhead, but it suffers from very high hardware costs and energy consumption caused by time-delay circuits. In this paper, we propose a near-field two dimension (2D) hierarchical beam training scheme to reduce the overhead without the need for extra hardware circuits. Specifically, we first formulate the multi-resolution near-field codewords design problem covering different angle and distance coverages. Next, inspired by phase retrieval problems in digital holography imaging technology, we propose a Gerchberg-Saxton (GS)-based algorithm to acquire the theoretical codeword by considering the ideal fully digital architecture. Based on the theoretical codeword, an alternating optimization algorithm is then proposed to acquire the practical codeword by considering the hybrid digital-analog architecture. Finally, with the help of multi-resolution codebooks, we propose a near-field 2D hierarchical beam training scheme to significantly reduce the training overhead, which is verified by extensive simulation results.

	\end{abstract}
	
	
	\section{Introduction}\label{S1}
	
With the emergence of new applications such as digital twins, 6G is expected to achieve a 10-fold increase in spectrum efficiency than 5G~\cite{6G,Ray_6G}. 
The extremely large-scale MIMO (XL-MIMO) is a promising technique for 6G to achieve ultra-high spectrum efficiency~\cite{XLMIMO,Near_field_Mag}. In XL-MIMO systems, the base station (BS) usually deploys an extremely large-scale antenna array (ELAA), which consists of hundreds or even thousands of antennas. ELAA in the XL-MIMO system is expected to drastically improve spatial resolution to realize a high spatial multiplexing gain in 6G.
In order to obtain spatial multiplexing gain, 
XL-MIMO should generate a directional 
beam with high array gain by beamforming. 
To support beamforming, beam training should be conducted to search the optimal beamforming vector, i.e., codeword, in the predefined codebook. As the number of BS antennas in XL-MIMO systems is much larger than that of 5G systems,
the high-dimensional XL-MIMO beam training overhead will be overwhelming.


	\subsection{Prior Works}
	There are two typical categories of beam training methods for MIMO, which are far-field beam training and near-field beam training respectively. For the first category, since the antenna number at BS is usually not very large in 3G-5G systems, the MIMO channel is modeled in the far-field region with the planar wave assumption, where the array response vector of the far-field channel is only related to the angle. In this case, the orthogonal Discrete Fourier Transform (DFT) codebook can be utilized in beam training to capture the physical angle information in the angle-domain of the channel paths~\cite{far_ce,far_ce2}. 
	However, since the size of the DFT codebook is proportional to the number of antennas at BS, the DFT codebook suffers from very high training overhead when it comes to XL-MIMO systems. Thus, to reduce the beam training overhead, some hierarchical beam training schemes were proposed~\cite{alkhateeb2014channel,David15}. The basic idea of the beam training is to search from the lowest-resolution codebook to the highest-resolution codebook layer by layer, where the angle range needed to be scanned reduces layer by layer gradually. With the help of hierarchical beam training, the overhead becomes proportional to the logarithm of the antenna number at BS~\cite{David15}.  
	
	As the antenna number dramatically increases in 6G XL-MIMO systems, the near-field range expands by orders of magnitude, which can extend to several hundred meters~\cite{Marzetta22}. Thus, the XL-MIMO channel should be modeled in the near-field region subjected to the spherical wave assumption. 
	In this case, the existing far-field beam training schemes may not be valid for the near-field XL-MIMO channel. To cope with this problem, near-field beam training should be utilized to match the near-field XL-MIMO channel feature.
 	For the second category, i.e., near-field beam training, the array response vector of the near-field channel is not only related to the angle but also to distance. 
 	Thus, to capture the physical angle information as well as distance information of the channel paths, a polar-domain codebook~\cite{Mingyao} should be utilized instead of a DFT codebook. Accordingly, the size of the polar-domain codebook is the product of the antenna number at BS and the number of sampled distance grids. Since only one angle and one distance can be measured in each time slot, the exhaustive search method for near-field beam training has a very high overhead~\cite{XiuCC}. To address this problem, we have proposed a fast time-delay based near-field beam training for XL-MIMO with low overhead~\cite{Rainbow}, where each antenna requires time-delay circuits to provide frequency-dependent phase shift. In specific, due to the near-field beam split effect in a wideband situation, near-field beams can be flexibly controlled by extra time-delay hardware circuits and then focus on different angles and distances at different frequencies in one time slot. However, the time-delay based beamforming structure will lead to not only high hardware costs but also very high energy consumption, especially for XL-MIMO systems with a large number of antennas.

	
	
	
	\subsection{Contributions}
	
	Thus, to design a general and low-overhead beam training scheme, we propose a near-field two dimension (2D) hierarchical beam training scheme by designing the multi-resolution codebooks referring to the hierarchical beam training in the far-field scenario. 
	Our contributions are summarized as follows.
	
	\begin{enumerate}
		
		\item We first formulate the problem of near-field codeword design. Specifically, compared with the far-field case, the ideal beam pattern of near-field codeword should not only cover a certain angle range but also a certain distance range. By considering ideal fully digital architecture, we provide the design problem of the near-field theoretical codeword. Then, based on the theoretical codeword, we formulate the problem of a practical codeword with assumptions of the hybrid digital-analog structure and quantized phase shifts in practice.
		
		\item In order to design the near-field theoretical codeword, inspired by the Gerchberg–Saxton (GS) algorithm in phase retrieval problems for digital holography imaging, we propose a GS-based theoretical codeword design algorithm for a fully digital architecture. Different from the original GS algorithm, we modify the transformation methods from Fourier transform to polar-domain transform to match the near-field assumption. Additionally, the power constraint instead of amplitude measurements are considered in each iteration to control the power of the codeword.

		\item Since fully digital architecture with high energy assumption is not available in a practical XL-MIMO system, we then design the practical codeword considering the hybrid digital-analog architecture.  
		Based on the theoretical codeword, an alternating optimization algorithm is proposed to acquire the practical codeword, where the digital beamforming vector and the analog beamforming matrix are optimized iteratively. Specifically, in each iteration, the digital beamforming vector is obtained by a closed-form solution. Meanwhile, phases of the entries in the analog beamforming matrix are solved individually by a high-efficient iterative search method.

		\item Next, we generate multi-resolution codebooks based on the practical codewords obtained by the alternating optimization algorithm. With the aid of multi-resolution codebooks with different angle coverages and distance coverages, we propose a near-field two dimension (2D) hierarchical beam training scheme. Specifically, codewords are searched in multi-resolution codebooks layer by layer, where angle and distance ranges are reduced gradually. Moreover, we provide the analysis of the proposed beam training overhead, which is proportional to the sum of the logarithm of the antenna number and the sampled distance grid number. Simulation results show that the proposed beam training scheme can reach sub-optimal achievable rate performance with low overhead.


	\end{enumerate}
	
	\subsection{Organization and Notations}
	{\it Organization}: The rest of the paper is organized as follows. In Section II, we first introduce the signal model, the near-field channel model, and the formulation of the near-field codebook design problem. In Section III, we provide the design of the theoretical codeword by the proposed Gerchberg-Saxton algorithm considering fully digital architecture. In Section IV, we propose an alternating optimization scheme to design the practical codeword with hybrid digital-analog architecture. Then, the proposed near-field 2D hierarchical beam training scheme is described in Section V. Simulation results and conclusions are provided in Section VI and Section VII, respectively.

	{\it Notations}: Lower-case and upper-case boldface letters ${\bf{a}}$ and ${\bf{A}}$ denote a vector and a matrix, respectively; ${{{\bf{a}}^H}}$ and ${{{\bf{A}}^{H}}}$ denote the conjugate transpose of vector $\bf{a}$ and matrix $\bf{A}$, respectively; ${{\|{\bf{a}}\|_2}}$ denotes the $l_2$ norm of vector $\bf{a}$; ${{\|{\bf{a}}\|_F}}$ denotes the Frobenius norm of vector $\bf{a}$. $ {\bf 0}_{N\times M} $ denotes $ N\times M $-dimensional null matrix. 
	Finally, $\cal CN\left(\boldsymbol{\mu},\bf{\Sigma} \right)$ denotes the probability density function of complex multivariate Gaussian distribution with
	mean $ \boldsymbol{\mu} $ and variance $ \bf{\Sigma} $.
	${{\cal U}(-a,a)}$ denotes the probability density function of uniform distribution on $(-a,a)$.
	
	\vspace{5mm}
	\section{System Model}\label{S2}
	In this section, we will first introduce the signal model of the XL-MIMO system. Then, the existing near-field channel model will be briefly reviewed. Finally, we formulate the problem of codeword design in the near-field scenario.
	
	\vspace{0mm}
	\subsection{Signal Model}\label{S2.1}
	We consider the scenario where the BS employs a $N$-element ELAA to communicate with a single-antenna user. Let ${\bf{h}}^H\in\mathbb{C}^{1\times N}$ denote the channel from the BS to the user. Since the XL-MIMO channel $ {\bf{h}}^H$ is generally dominated by a few main paths, we only need to search the physical location of the main paths by beam training instead of acquiring the explicit channel information~\cite{Liye,Min}. Therefore, the main path is concerned in this paper, and the corresponding beam training method will be investigated to search for the optimal beamforming vector to align with the main path. 
	
	Take downlink transmission as example, the received signal $ y $ can be represented by
	\begin{equation}\label{eq1}
	{{y}} = {\bf{h}}^H{\bf{v}}s + {{n}},
	\end{equation}
	where ${{\bf{v}}}\in\mathbb{C}^{{N\times 1}}$ represents the beamforming vector at the BS, which is
	essentially a codeword chosen from the predefined codebook, $ s $ represents the symbol transmitted by the BS, and ${{\bf{n}}}\sim{\cal C}{\cal N}\left( {{\bf{0}}_{N},\sigma^2{\bf{I}}_{N}} \right)$ represents the received noise with ${\sigma^2}$ representing the noise power. The beam training is to measure the power of $ y $ to find the best codeword from the codebook.

	Next, we will briefly review the existing near-field XL-MIMO channel model for existing near-field beam training schemes.

	\vspace{0mm}
	\subsection{Near-Field XL-MIMO Channel Model}\label{S2.2}
	
	

	When the distance between the BS and the UE is smaller than the Rayleigh distance~\cite{NearFar}, the near-field XL-MIMO channel should be modeled with the spherical wave assumption, which can be expressed by
	\begin{equation}\label{eq6}
	{\bf{h}}=\sqrt{{N}}{\alpha}{\bf{b}}\left({\theta},r\right).
	\end{equation}
	where $ \alpha $ is the complex path gain. ${\bf{b}}\left({\theta},r\right)$ represents the near-field array response vector, which can be represented by~\cite{Mingyao}
	\begin{equation}\label{eq7}
	{\bf{b}}(\theta, r) = \frac{1}{\sqrt{N}}[e^{-j{\frac{2\pi}{\lambda}}(r^{(1)} - r)},\cdots, e^{-j{\frac{2\pi}{\lambda}}(r^{(N)} - r)}]^H,
	\end{equation}
	where $r$ represents the distance from the UE to the center of the antenna array, 
	
	\hspace{-4mm}$r^{(n)} = \sqrt{r^2 + \delta_n^2d^2 - 2r\delta_n d\theta}$ represents the distance from the UE to the $n$th BS antenna, and $\delta_n = \frac{2n - N - 1}{2}$ with $n = 1,2,\cdots, N$. 
	
	Before data transmission, beam training should be applied to estimate the physical angles and distances of near-field channel paths. The near-field response vector $ {\bf{b}}(\theta, r) $  implies that the optimal beam training codeword should focus on the spatial angle $ \theta $ and BS-UE distance $ r $. The existing near-field beam training scheme is conducting an exhaustive search in the polar-domain codebook~\cite{Mingyao}, which can be represented as  
	\begin{equation}\label{eq8}
	\begin{aligned}
	{\bf{A}}=[{\bf{b}}(\theta_1, r_1^1),\cdots,{\bf{b}}(\theta_1, r_{1}^{S_{1}}),\cdots,{\bf{b}}(\theta_{N}, r_{N}^{S_N})],
	\end{aligned}
	\end{equation}
	where each column of polar-domain codebook $\bf{A}$ is a codeword aligned with the grid  ($\theta_{n}$, $r_{{n}}^{s_{n}}$), 
	with $s_{n}=1,2,\cdots,{S_{n}}$, $S_{n}$ denotes the number of sampled distance grids at $\theta_{{n}}$. Therefore, the number of total sampled grids of the whole propagation environment is $S=\sum\limits_{{n} = 1}^{N}S_{n}$. Apparently, in XL-MIMO systems, the codebook should not only sample angle but also distance, which leads to a large-size codebook and unfordable beam training overhead.
	Thus, to address this problem, we design the hierarchical near-field codebook with multi-resolution codebooks, and then propose the corresponding near-field 2D hierarchical beam training. To design the multi-resolution near-field codebooks, we will first formulate the design problem of a near-field codeword with different angle coverage and distance coverage. 
	
	\vspace{0mm} 
	\subsection{Formulation of Codebook Design Problem}\label{S3.1}
	Suppose the angle coverage and distance coverage of codeword $ \bf{v} $ are $ {\bf{B}}_{\mathbf{v},\theta} \triangleq [\theta, \theta+B_{\theta}]  $ and $ {\bf{B}}_{\mathbf{v},r} \triangleq [r, r+B_{r}]  $, where $ B_{\theta} $ and $ B_{r} $ are the angle sampled step and distance sampled step. The ideal beam pattern vector of the codeword $ \bf v $ is denote as 
	\begin{equation}\label{ideal_bp}
	{\bf g}_{\bf{v}} = \left[ {{g}}_{\bf{v}}(\theta_1, r_1^1),\cdots, {{g}}_{\bf{v}}(\theta_{N}, r_{N}^1),\cdots,{{g}}_{\bf{v}}(\theta_{N}, r_{N}^{S_N}) \right] ,
	\end{equation} 
	where $  g_{\bf{v}}(\theta ,r) =  \left| g_{\bf{v}}(\theta ,r)\right|e^{jf_{\bf v}(\theta ,r)}  $ is the theoretical beamforming gain.  
	The amplitude information $ \left| g_{\bf{v}}(\theta ,r)\right|  $  of the ideal beam pattern can be further represented by 
	\begin{equation}\label{coverage}
	\left| g_{\bf{v}}(\theta ,r)\right| = \begin{cases} \sqrt{C_{\mathbf{v}}}, & \theta \in {\bf B}_{\mathbf{v},\theta},\  r \in {\bf B}_{\mathbf{v},r} \\ {0}, & \theta \notin {\bf B}_{\mathbf{v},\theta}, \  r \notin {\bf B}_{\mathbf{v},r} \end{cases}.
	\end{equation}
	For the ideal beam pattern in~(\ref{ideal_bp}), the amplitude information $ \left| g_{\bf{v}}(\theta ,r)\right|  $ of ideal beam pattern vector in target angle coverage and distance coverage are fixed and flattened while other beamforming gains are zero. Meanwhile, the phase information $ f_{\bf v}(\theta ,r) $ of the ideal beam pattern vector can be designed flexibly. Compared to a far-field codeword, the near-field codeword should cover not only a certain angle range but also a certain distance range. 
	
	To evaluate the effectiveness of the codeword $ \bf{v} $, we reference $ G\left({\bf{v}}, \theta ,r\right)  $ as the beamforming gain of $ \bf{v} $ in the angle $ \theta $ and the distance $ r $. 
	The $ G\left( {\bf{v}}, \theta ,r\right)  $  can be represented as 
	\begin{equation}
	G({\bf{v}}, \theta ,r)=\sqrt{N} \mathbf{b}(\theta ,r)^H \mathbf{v}.
	\end{equation}
	Thus, according to the definition of polar-domain codebook $\bf A$ in (\ref{eq8}), the beam pattern obtained by beamforming with codeword $ \bf{v} $ can be presented as $ {\bf{A}}^H {\bf{v}} $.
	
	The aim of designing a codeword is to make the beam
	pattern ${\bf{A}}^H {\bf{v}} $ obtained by beamforming with the codeword $ {\bf{v}} $ as close as possible to the ideal beam pattern $ \bf{g}_{\bf v} $. Thus, the objective of the theoretical codeword $ \bf{v} $ design can be express as 
	\begin{equation}\label{P1}
	\min _{{\bf{v}}, f(\theta ,r)}\left\|{\bf{A}}^H {\bf{v}}-\bf{g}_{\bf v}\right\|_2^2. \tag{P1}
	\end{equation}
	
	In (P1), the ideal theoretical codeword $ \bf v $ can only be realized by the fully digital architecture, where each antenna requires one dedicated radio frequency (RF) chain to realize fully digital signal processing. However, fully digital architecture in the XL-MIMO system results in unaffordable energy consumption. In fact, a hybrid digital-analog structure is usually preferred in XL-MIMO systems to improve energy efficiency~\cite{Huang}. In this structure, we need to design practical codewords considering the hardware constraints in terms of phase shifter resolution and the number of radio frequency (RF) chains $ N_{\rm RF} $~\cite{Gen17}. 
	
	Specifically, based on the ideal theoretical codeword $ \bf{v} $, the design of the practical codeword $ \bf{v}_{\rm{p}} \triangleq \bf{F}_{\rm{RF}} \bf{f}_{\rm{BB}} $ can be represented as 
	\begin{equation}
	\begin{aligned}
	\min_{{\bf{F}}_{\rm{RF}}, \bf{f}_{\rm{BB}}}
	&\left\|\bf{v}-{\bf{F}}_{\rm{RF}} {\bf{f}}_{\rm{BB}}\right\|_2 \\
	\text { s.t. } &\left\|\bf{F}_{\rm{RF}} {\bf{f}}_{\rm{BB}}\right\|_2=1, \\
	& \left[\bf{F}_{\rm{RF}}\right]_{n, i}=e^{j \delta_{n,i}},\delta_{n,i} \in \Phi_b\\
	& n=1,2, \ldots, N, i=1,2, \ldots, N_{\mathrm{RF}},
	\end{aligned}\tag{P2}
	\end{equation}
	where the $ {\bf{F}}_{\rm{RF}} \in \mathbb{C}^{N\times N_{\rm RF}} $  and $  {\bf{f}}_{\rm{BB}} \in \mathbb{C}^{ N_{\rm RF} \times 1} $ are the analog beamforming matrix and the digital beamforming vector. $  \boldsymbol{\Phi}_b=\left[\pi\left(-1+\frac{1}{2^b}\right), \pi\left(-1+\frac{3}{2^b}\right), \ldots \pi\left(1-\frac{1}{2^b}\right)\right] $ is the set of quantized phase shifters with $ b $ bits.
	
	
	All the codewords in the codebook can be designed based on (P1) and (P2). Next, we introduce the design method of the theoretical codeword $ \bf v $ in Section \ref{S3} and practical codeword $ {\bf v}_{\rm p} $ Section \ref{S4}.

	\vspace{5mm}
	\section{Proposed Gerchberg-Saxton Algorithm based Near-Field Theoretical Codeword Design}\label{S3}
	In this section, we will first briefly review the Gerchberg-Saxton algorithm applied in the phase retrial problem in the hologram optical system, and the relationship between the phase retrieval problem and the codeword design problem is analyzed. Next, we propose a GS-based theoretical codeword design scheme. Finally, the convergence property of the GS algorithm in near-field codeword design is provided. 
	\vspace{-0mm}

	\subsection{Preview of the Phase Retrieval Problem and Gerchberg-Saxton algorithm}\label{S3.2}
	
	\subsubsection{Phase retrieval problem in digital holography imaging}
	
In recent years, with the development of modern optics and computer science, digital holography imaging technology has changed the traditional imaging object-image relationship and structure by combining the front-end optical system design with the back-end signal processing. The back-end signal processing algorithm of the original data collected by the camera can break through the traditional imaging bottleneck.
	
	In specific, in optical systems, the amplitude information is easy to measure, while the direct recording of the phase information is not allowed. The reason is that the electromagnetic field oscillates at a very high frequency that rare electronic measurement devices can follow~\cite{phase_retrival}. Thus, in order to realize the imaging of the original object, one of the most important problems in digital holography imaging technology is conducting phase retrieval~\cite{holographic}. Fortunately, with the help of the measured amplitude information, some signal processing algorithms offer alternative methods for recovering the phase information of optical images without requiring sophisticated devices. 
	
	Reviewing the theoretical codeword design problem in (\ref{P1}), it is obvious that the problem (\ref{P1}) is similar to the phase retrieval in digital holography imaging, where the phase information ($ f_{\bf v}(\theta ,r) $ of the ideal beam pattern vector) should be obtained by measured amplitude information ($ \left| g_{\bf{v}}(\theta ,r)\right|  $ of ideal beam pattern vector).



\begin{figure}
	\centering
	\subfigure[Illustration of GS algorithm in iterative phase retrieval problem.]{
		\begin{minipage}[t]{0.5\linewidth}
			\centering			
			\hspace*{-3.5mm}\includegraphics[width=3.3in]{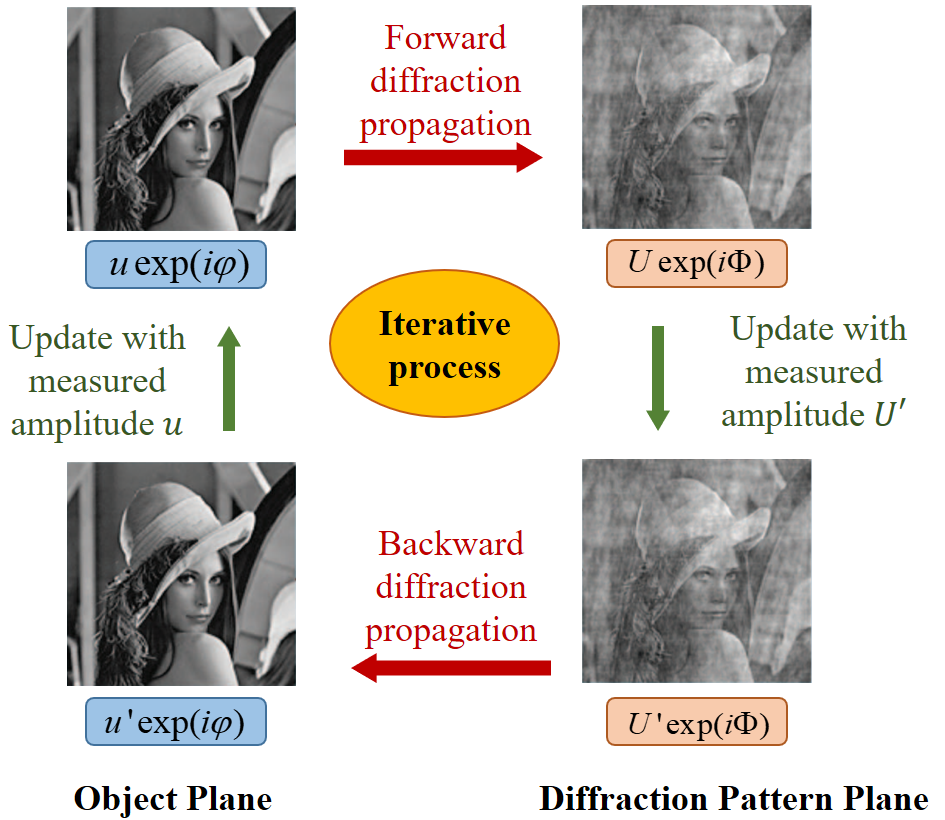}
		\end{minipage}
	}%
	\subfigure[Illustration of GS algorithm in codeword design]{
		\begin{minipage}[t]{0.5\linewidth}
			\centering
			\hspace*{-3.5mm}\includegraphics[width=3.3in]{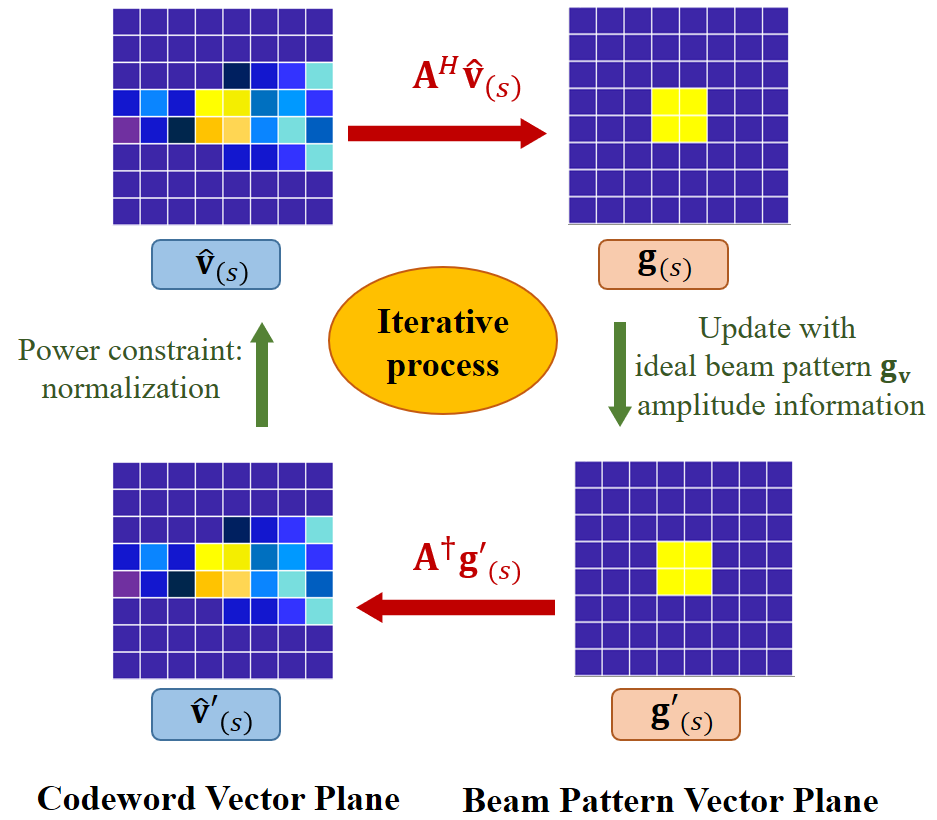}
		\end{minipage}
	}%
	\centering
	\vspace{-0mm}
	\caption{Comparisons of the original and improved GS algorithm.}\vspace{-0mm}
	\label{GS}
\end{figure}

	\subsubsection{Gerchberg-Saxton algorithm}
	One of the most popular methods to solve the phase retrieval problem is Gerchberg–Saxton (GS)-based algorithm~\cite{GS_Ori,GS0} as shown in Fig.~\ref{GS} (a), where two amplitude measurements are iteratively imposed in the object plane and diffraction pattern plane~\cite{Caolc,GS1}. It is worth noting that the diffraction pattern plane is also known as the Fourier plane since the complex-valued wavefronts in the object and the diffraction pattern planes are usually connected through a Fourier transform with each other. 
	
	Specifically, the GS algorithm initializes in the object plane, where the initial complex-valued wavefronts are created by combining the measured amplitude information with the random phase information. The iteration process of the GS algorithm consists of four steps: i) The forward diffraction propagation of the wavefronts in the object plane provides complex-valued wavefronts in the diffraction pattern plane; ii) Update the complex-valued wavefronts in the diffraction pattern plane: the amplitude information is substituted with the measured amplitude information $ U' $; iii) The backward diffraction propagation provides the complex-valued wavefronts in the object plane; iv) Update the complex-valued wavefronts in the diffraction plane: The amplitude information in the object plane is substituted with the measured amplitude information. The result of the GS algorithm is the recovered complex-valued wavefronts in the diffraction pattern plane. 
	
	Some modified versions of the GS algorithm have been proposed afterward~\cite{GS_After} to match various imaging problems. Instead of utilizing the GS algorithm in the imaging problem, we improved the GS algorithm in the near-field codeword design problem. In specific, we replace one of updating processes with measured amplitude information by applying normalization to match the power constraint of the codeword.
	
	\vspace{-0mm}
	\subsection{Design of the Theoretical Codeword $ \bf v $}\label{S3.3}
	In order to solve the (P1), we draw the experience from the Gerchberg–Saxton (GS) algorithm, which is widely applied in phase retrieval problems in digital hologram imaging of optical systems. In the phase retrieval problem, the phase information needed to be obtained with the fixed amplitude information, which is the same as the phase information $ f_{\bf v}(\theta,r) $ design of the ideal beam pattern in the problem (\ref{P1}). Specifically, the proposed GS-based near-field codeword design procedure is shown in \textbf{Algorithm 1}.
	
	\begin{algorithm}[htbp]
		\caption{GS-based theoretical codeword design}
		\textbf{Inputs}: $|{\bf g}_{\bf{v}}|$, ${C}_{\bf{v}}$, ${S}_{max}$, ${\bf A}$, $B_{\mathbf{v},\theta}$, $B_{\mathbf{v},r}$.
		\\\textbf{Initialization}: randomly generate $ f_{(0)}\left( \theta,r\right)  $ and obtain the ${\bf g}_{(0)}$ by (\ref{G0}).
		\\1. $ \hat{\bf v}'_{(0)} = \left({\bf A} {\bf A}^H\right)^{-1} {\bf A} {\bf g}_{(0)}   $ 
		\\2. Obtain $ \hat{\bf v}_{(1)} $ by normalizing $ \hat{\bf v}'_{(0)} $
		\\3. \hspace*{+3mm}\textbf{for} $s = 1,2,\cdots, S_{\rm max}$ \textbf{do}
		\\4. \hspace*{+3mm}\hspace*{+3mm}calculate $ {\bf g}_{(s)}  $ based on $ \hat{\bf v}_{(s)} $ by (\ref{v1})  
		\vspace*{0.6mm}
		\\5. \hspace*{+3mm}\hspace*{+3mm}calculate $ {\bf g}'_{(s)}  $ based on $ {\bf g}_{(s)}  $ and $ {\bf g_v}  $ by (\ref{G}) 
		\vspace*{0.8mm}
		\\6. \hspace*{+3mm}\hspace*{+3mm}calculate $ \hat{\bf v}'_{(s)}  $ based on $ {\bf g}'_{(s)}  $ by (\ref{v2})  
		\vspace*{0.8mm}
		\\7. \hspace*{+3mm}\hspace*{+3mm}\textbf{if} $ s < S_{max}$
		\\8. \hspace*{+3mm}\hspace*{+3mm}\hspace*{+3mm}calculate $ \hat{\bf v}_{(s+1)} $ based on $ \hat{\bf v}'_{(s)}  $ by (\ref{v_nor}) 	
		\\9. \hspace*{+3mm}\hspace*{+3mm}\textbf{end if}
		\\10. \hspace*{+3mm}\hspace*{+0mm}\textbf{end for}
		\vspace*{0.5mm}
		\\11. \hspace*{0mm}\hspace*{+0mm}$ {\bf v }= \hat{\bf v}'_{(S_{max})}/ ||{\hat{\bf v}}'_{(S_{max})}||_{2}$ 
		\vspace*{0.6mm}
		\\\textbf{Output}: Theoretical codeword $ \bf v $.
	\end{algorithm}
	
	
	For notation simplicity, in the description of the GS algorithm, we use $ \hat{\bf v}_{(s)} $, 
	${ \bf g}_{(s)} $, 
	${ \bf g}'_{(s)}$, and ${\hat{\bf v}}'_{(s)} $ to denote the designed codeword vector, the beam pattern vector realized by the designed codeword, the revised beam pattern vector with ideal beam pattern amplitude, and the codeword vector obtained by revised beam pattern vector in the $ s $-th iteration of GS algorithm. 
	
	Before the GS algorithm starts, we should first obtain the initial beam pattern vector $ {\bf g}_{(0)} $ with randomly generated phase $ f_{(0)}\left( \theta,r\right)  $ and amplitude information $ {g}_{\bf{v}}(\theta, r) $ of ideal beam pattern vector ${\bf g}_{\bf{v}}$. In this way, the $ {\bf g}_{(0)} $ can be represented as  
	\begin{equation}\label{G0}
	\begin{aligned}
	{\bf g}_{(0)} \!\!= \! \Big[ & \left|{g}_{\bf{v}}(\theta_1, r_1^1)\right|\!{f_{(0)}\!(\theta_1, r_1^1)},\!\cdots\!,\!|{{g}}_{\bf{v}}(\theta_{N}, r_{N}^1)|{f_{(0)}\!(\theta_{N}, r_{N}^1)},
	&\cdots,|{{g}}_{\bf{v}}(\theta_{N}, r_{N}^{S_N})|{f_{(0)}(\theta_{N}, r_{N}^{S_N})} \Big].
	\end{aligned}
	\end{equation}
	In $s$-th iteration, with provided designed $ \hat{\bf v}_{(s)} $, 
	\begin{equation}\label{v1}
	{\bf g}_{(s)} =  {\bf A}^H \hat{\bf v}_{(s)}.
	\end{equation}
	Then, in order to maintain the amplitude information of the ideal beam pattern vector $ {\bf g}_{\bf v}$ to approach the ideal beam pattern, we assign the amplitude information $ |{\bf g_v}(\theta,r)| $ of ideal beam pattern $ {\bf g_v} $ to $ {\bf g'}_{(s)} $, and the phase information $ {f_{\hat{\bf v}_{(s)}}(\theta,r)} $ of current beam pattern $ {\bf g}_{(s)} $ to $ {\bf g'}_{(s)} $.
	In this case, the  $ {\bf g'}_{(s)} $ can be presented as
	\begin{equation}\label{G}
	\begin{aligned}
	{\bf g'}_{(s)} \!\!= \! \Big[ & \left|{g}_{\bf{v}}(\theta_1, r_1^1)\right|\!{f_{\hat{\bf v}_{(s)}}\!(\theta_1, r_1^1)},\cdots,|{{g}}_{\bf{v}}(\theta_{N}, r_{N}^1)|{f_{\hat{\bf v}_{(s)}}(\theta_{N}, r_{N}^1)},
	&\cdots,|{{g}}_{\bf{v}}(\theta_{N}, r_{N}^{S_N})|{f_{\hat{\bf v}_{(s)}}(\theta_{N}, r_{N}^{S_N})} \Big].
	\end{aligned}
	\end{equation}
	Base on the (P1), given $ {\bf g'}_{(s)}  $, the $ {\hat{\bf v}}'_{(s)}  $ can be obtained by least square algorithm as 
	\begin{equation}\label{v2}
	{\hat{\bf v}}'_{(s)} = \left({\bf A} {\bf A}^H\right)^{-1} {\bf A} {\bf g}_{(s)}' = {\bf A}^{\dagger} {\bf g}_{(s)}' ,
	\end{equation}
	where the pseudo inverse of $ {\bf A}^H $ is denoted as $ {\bf A}^{\dagger} $.
	Finally, we normalize the $ {\hat{\bf v}}'_{(s)} $ as
	\begin{equation}\label{v_nor}
	{\hat{\bf v}}_{(s+1)} = {\hat{\bf v}}'_{(s)}/ ||{\hat{\bf v}}'_{(s)}||_{2}.
	\end{equation}
	After the iteration number reaches $ S_{max} $, we utilize $ \hat{\bf v}'_{(S_{max})} $ to obtain the designed theoretical codeword $ {\bf v} $.

	\subsection{Convergence Property of GS Algorithm in Near-Field Codeword Design}
	
	As mentioned before, the original GS algorithm assumes that the object and the diffraction pattern planes are connected through a Fourier Transform (FT). 
	The convergence of the original GS algorithm with FT assumption is proved based on Parseval's theorem of FT~\cite{Parseval}, where the energy of wavefronts in the object and the diffraction pattern planes before and after FT and inverse FT are the same. However, the codeword vector plane and beam pattern vector plane in the proposed GS algorithm are connected with the polar-domain transformation, which does not satisfy Parseval's theorem. Thus, the convergence property of the proposed GS algorithm based on polar-domain transformation in near-field codeword design should be analyzed.
	
	In this paper, the convergence of the proposed GS algorithm is supervised by the squared error in each iteration. 
	Specifically, the squared error of the beam pattern plane in $ s $-th iteration can be presented as 
	\begin{equation}\label{proof1}
	\begin{aligned}
	{E_{(s)}}&=\iint\|{\bf g}_{(s)}(\theta, r)-{\bf {g}'}_{(s)}(\theta, r)\|_2^2 d \theta d r.\\
	&=\iint\|{\bf A}^{H}{\hat{\bf v}_{(s)}(u,w) }-{\bf A}^{H}{{\hat{\bf v}'_{(s)}(u,w) }}\|_2^2 d u d w\\
	\end{aligned}
	\end{equation}
	It is worth noting that the codewords in the polar-domain codebook $ {\bf A}^{H} $ have been rearranged, where the codewords aligned with the largest distance $ {S_{n}} $ of each $ \theta_{n} $ are brought to the front columns of $ {\bf A}^{H} $. Thus, the $ {\bf A}^{H} $ can be rewritten as
	\begin{equation}
	\begin{aligned}
	{\bf A}^{H} = [{\bf A}_1,{\bf A}_2]^H,
	\end{aligned}
	\end{equation}
	where $ {\bf A}_1 = [{\bf{b}}(\theta_1, r_{1}^{S_{1}}),{\bf{b}}(\theta_2, r_{2}^{S_{2}}),\cdots,{\bf{b}}(\theta_{N}, r_{N}^{S_N})] $,\vspace{0.25cm} 
	
	\hspace{-4mm} $ {\bf{A}}_2=[{\bf{b}}(\theta_1, r_1^1),\!\cdots,\!{\bf{b}}(\theta_1, r_{1}^{S_{1}-1}),\!\cdots,\!{\bf{b}}(\theta_{N}, r_{N}^{1}),\!\cdots,\!{\bf{b}}(\theta_{N}, r_{N}^{{S_N}-1})] $.\vspace{0.25cm}
	Since the $ {S_{n}} $ in each column $ {\bf{b}}(\theta_n, r_{1}^{S_{n}}) $ of $ {\bf A}_1 $ is larger than Rayleigh distance, $ {\bf{b}}(\theta_n, r_{1}^{S_{n}}) $ approximates to the far-field codeword aligned with the physical direction $ \theta_n $. In this case, the $ {\bf A}_1 $ is equal to a far-field DFT codebook. Thus, $ {\bf A}_1^{H}\left( {\hat{\bf v}_{(s)}(u,w)\!-\!{\hat{\bf v}'_{(s)}(u,w) }}\right)  $ is an FT process, which  satisfies Parseval's theorem as 
	\begin{equation}\label{proof9}
	\begin{aligned}
	\iint\|\left( {\hat{\bf v}_{(s)}(u,w)\!-\!{\hat{\bf v}'_{(s)}(u,w) }}\right) \|^2_2 d u d w \\
	= \!\!\iint\|{\bf A}_1^{H}\left( {\hat{\bf v}_{(s)}(u,w)\!-\!{\hat{\bf v}'_{(s)}(u,w) }}\right) \|^2_2 d u d w.
	\end{aligned}
	\end{equation}
	
	Therefore, $ {E_{(s)}} $ can be further expressed as 
	\begin{equation}\label{proof8}
	\begin{aligned}
	{E_{(s)}}
	=& \!\!\iint\|{\bf A}_1^{H}\left( {\hat{\bf v}_{(s)}(u,w)\!-\!{\hat{\bf v}'_{(s)}(u,w) }}\right) \|^2_2 \\
	&+\|{\bf A}_2^{H}\left( {\hat{\bf v}_{(s)}(u,w)\!-\!{\hat{\bf v}'_{(s)}(u,w) }}\right) \|^2_2d u d w.\\
	\geq&\!\!\iint\|\left( {\hat{\bf v}_{(s)}(u,w)\!-\!{\hat{\bf v}'_{(s)}(u,w) }}\right) \|^2_2 
	\end{aligned}
	\end{equation}

	The squared error of the codeword vector plane of $s+1$-th iteration for the GS algorithm can be expressed as 
	\begin{equation}\label{E0}
	\begin{aligned}
	{E_{(s)}^{0}}&=\iint\|{\hat{\bf v}}_{(s+1)}(u, w)-{\hat{\bf v}'}_{(s)}(u, w)\|_2^2 d u d w.
	\end{aligned}
	\end{equation}
	
	Then, we provide \textbf{Lemma 1} to show the change of squared error between adjacent iteration in codeword vector plane.
	
	\textbf{Lemma 1}: \textit{In the codeword vector plane of GS algorithm, the error between $ \hat{\bf v}_{(s)}(u,w) $ and $ {\hat{\bf v}'_{(s)}(u,w) } $ not less than than the error between $ \hat{\bf v}_{(s+1)}(u,w) $ and $ {\hat{\bf v}'_{(s)}(u,w) } $, i.e.,  $ \| {\hat{\bf v}_{(s)}(u,w)\!-\!{\hat{\bf v}'_{(s)}(u,w) }} \|^{2}_{2} > \| {\hat{\bf v}_{(s+1)}(u,w)\!-\!{\hat{\bf v}'_{(s)}(u,w) }} \|^{2}_{2} $}.
	
	\textit{proof:} See Appendix A.
	
	From the (\ref{proof8}), (\ref{E0}), and \textbf{Lemma 1}, we can derive that
	\begin{equation}\label{eq2}
	\begin{aligned}
	{E_{(s)}^{0}} \leq \iint\|{\hat{\bf v}}_{(s)}(u, w)-{\hat{\bf v}'}_{(s)}(u, w)\|_2^2 d u d w \leq {E_{(s)}}
	\end{aligned}
	\end{equation}
	
	On the other hand, $ {E_{(s)}^{0}} $ can be further expressed as
	\begin{equation}\label{eq4}
	\begin{aligned}
	{E_{(s)}^{0}} &=\iint\|{\hat{\bf v}}_{(s+1)}(u, w)-{\hat{\bf v}'}_{(s)}(u, w)\|_2^2 d u d w\\
	& = \iint\|{\bf A}^{\dagger}{{\bf g}_{(s+1)}(\theta,r) }-{\bf A}^{\dagger}{{{\bf g}'_{(s)}(\theta,r) }}\|_2^2 d \theta d r.
	\end{aligned}
	\end{equation}
	Utilizing the uniqueness of pseudo inverses, we can easily know that $ {\bf A}^{\dagger} = \left[ ({\bf A}^{H})^{-1}, {\bf 0}_{(S-N )\times N} \right] $. In this case, since $ {\bf A}_1^{H}\left( {{\bf g}_{(s+1)}(\theta,r)\!-\!{{\bf g}'_{(s)}(\theta,r) }}\right)  $ is a inverse FT process, which also satisfies Parseval's theorem. Thus,
	\begin{equation}\label{eq5}
	\begin{aligned}
	{E_{(s)}^{0}} &= \iint\|{{\bf g}_{(s+1)}(\theta,r) }-{{{\bf g}'_{(s)}(\theta,r) }}\|_2^2 d \theta d r.
	\end{aligned}
	\end{equation}
	Similar to \textbf{Lemma 1}, we can obtain that 
	\begin{equation}\label{eq10}
	\begin{aligned}
	\|{{\bf g}_{(s+1)}(\theta,r) }\!-\!{{{\bf g}'_{(s)}(\theta,r) }}\|_2^2\! \geq \!\|{{\bf g}_{(s+1)}(\theta,r) }\!-\!{{{\bf g}'_{(s+1)}(\theta,r) }}\|_2^2
	\end{aligned}
	\end{equation}
	Thus, given (\ref{eq5}) and (\ref{eq10})
	\begin{equation}\label{eq3}
	\begin{aligned}
	{E_{(s)}^{0}}\geq\iint\|{{\bf g}_{(s+1)}(\theta,r) }-{{{\bf g}'_{(s+1)}(\theta,r) }}\|_2^2 d \theta d r = {E_{(s+1)}}.
	\end{aligned}
	\end{equation}

	Combining the equation (\ref{eq2}) and (\ref{eq3}), we can observe that 
	\begin{equation}\label{conc}
	\begin{aligned}
	{E_{(s+1)}} \leq {E_{(s)}^{0}} \leq {E_{(s)}},	
	\end{aligned}
	\end{equation}
	which means that the squared error in each iteration decreases. Thus, the convergence property of the proposed GS algorithm is proven.


	\vspace{0mm}
	
	\section{Proposed Alternating Optimization based Near-Field Practical Codeword Design}\label{S4}
    
    It is well known that each antenna requires one dedicated radio-frequency (RF) chain to realize the fully digital architecture. In this way, an XL-MIMO system with a very large number of antennas leads to an equally large number of RF chains, which will result in unaffordable hardware costs and energy consumption. To solve this problem, hybrid digital-analog architecture is preferred in practice, where the fully digital beamforming matrix is decomposed into a high-dimensional analog beamforming matrix and a low-dimensional digital beamforming vector. Moreover, quantized phase shifts instead of continuous quantized phase shifts are accessible for realizing analog beamforming matrix. Thus, in this section, alternating optimization is proposed for practical codeword design considering the hybrid digital-analog architecture and quantized phase shifts.

	Based on the theoretical codeword $ \bf v $ obtained by \textbf{Algorithm 1}, we solve the practical codeword $ {\bf v}_{\rm p}  $ design problem (P2) by alternating optimizing the digital beamforming vector $ \bf{f}_{\rm{BB}} $ and the analog beamforming matrix $  {\bf{F}}_{\rm{RF}} $ considering the hardware constraints. \textbf{Algorithm 2} provides the specific procedure to design the practical codeword. 
	\vspace{0mm}
	\begin{algorithm}[htbp]
		\caption{Practical codeword design}
		\textbf{Inputs}: ${\bf{v}}$, ${\bf{T}}_{max}$, ${\bf{P}}_{max}$, ${\bf \Phi}_b$, $N$, $N_{\rm RF}$.
		\\\textbf{Initialization}: randomly generate $ {\bf F}_{\rm RF}^0 $.
		\\1. \textbf{for} $t = 1,2,\cdots, T_{\rm max}$ \textbf{do}
		\\ // Design the digital beamforming vector.
		\\2. \hspace*{+3mm}calculate the $ {\bf f}_{\rm BB}^{t}  $ by  (\ref{FBB})  
		\\ // Design the analog beamforming matrix.
		\\3. \hspace*{+3mm}\textbf{for} $p = 1,2,\cdots,P_{max}$ \textbf{do}
		\\4. \hspace*{+6mm}\textbf{for} $ n = 1,2,\cdots,N$ \textbf{do}
		\\5. \hspace*{+9mm}\textbf{for} $i = 1,2,\cdots,N_{\rm RF}$ \textbf{do}
		\\6. \hspace*{+12mm}Search $ \delta_{n,i} $ to satisfy (\ref{P2_3})
		\\7. \hspace*{+9mm}\textbf{end for}
		\\8. \hspace*{+6mm}\textbf{end for}
		\\9. \hspace*{+6mm}\textbf{if} $ \delta_{n,i}^{p-1}= \delta_{n,i}^{p}$ \textbf{then}
		\\10.\hspace*{+9mm}Jump to Step2
		\\11.\hspace*{+6mm}\textbf{end if}
		\\12.\hspace*{+3mm}\textbf{end for}
		\\13.\hspace*{+3mm}obtain the $ {\bf F}_{\rm RF}^{t} $ by utilizing (\ref{FRF})
		\\14. \textbf{end for}
		\\\textbf{Output}: $ {\bf f}_{\rm BB}={\bf f}_{\rm BB}^{T_{max}}  $, $ {\bf F}_{\rm RF}={\bf F}_{\rm RF}^{T_{max}} $, $ {\bf v}_{p}={\bf F}_{\rm RF}^{T_{max}}{\bf f}_{\rm BB}^{T_{max}} $
	\end{algorithm}
	\vspace{0mm}
	
	For the given analog beamforming matrix $  {\bf{F}}_{\rm{RF}} $, the optimization problem of the digital beamforming vector $ \bf{f}_{\rm{BB}} $ can be expressed as 
	\begin{equation}
	\min _{\bf{f}_{\rm{BB}}}\left\|\bf{v}-\bf{F}_{\mathrm{RF}} \bf{f}_{\mathrm{BB}}\right\|_2,\tag{P2.1}
	\end{equation}
	which can be solved by least square as 
	\begin{equation}\label{FBB}
	\hat{\bf{f}}_{\mathrm{BB}}=\left({\bf{F}}_{\mathrm{RF}}^H \bf{F}_{\mathrm{RF}}\right)^{-1} {\bf{F}}_{\mathrm{RF}}^H \bf{v}
	\end{equation}
	
	Then, for the given analog beamforming vector $ \bf{f}_{\rm{BB}}  $, the optimization problem of $  {\bf{F}}_{\rm{RF}} $ can be expressed as 
	\begin{equation}
	\begin{aligned}
	\min_{{\bf{F}}_{\rm{RF}}}
	&\left\|\bf{v}-{\bf{F}}_{\rm{RF}} {\bf{f}}_{\rm{BB}}\right\|_2 \\
	\text { s.t. } &\left\|\bf{F}_{\rm{RF}} {\bf{f}}_{\rm{BB}}\right\|_2=1, \\
	& \left[\bf{F}_{\rm{RF}}\right]_{n, i}=e^{j \delta_{n,i}},\delta_{n,i} \in \Phi_b, \\
	& n=1,2, \ldots, N, i=1,2, \ldots, N_{\mathrm{RF}},
	\end{aligned}\tag{P2.2}
	\end{equation}
	The optimization of $ {\bf F}_{\rm RF} $ problem (P2.2) can be converted to the minimization absolute value of each entry of the vector $ \bf{v}-{\bf{F}}_{\rm{RF}} {\bf{f}}_{\rm{BB}} $. Hence, the problem (P2.2) can be transformed into $ N $ sub-problems, which can be optimized one by one. The $ n $-th sub-problem is rewritten as 
	\begin{equation}\label{P2_3}
	\begin{aligned}
	\min _{\theta_1, \theta_2, \ldots, \theta_{N_{\mathrm{RF}}}} & \left| \left[{\bf v} \right]_{n}  -\sum_{i=1}^{N_{\mathrm{RF}}}\left[\bf{f}_{\mathrm{BB}}\right]_i e^{j \delta_{n,i}}\right| \\
	\text { s.t. } & \delta_{n,i} \in \Phi_b, i=1,2, \ldots, N_{\mathrm{RF}} \text {. }
	\end{aligned}
	\end{equation}
	To obtain the solution to (\ref{P2_3}), the exhaustive search is a obvious choice, where all the combination of $ \delta_{n,1}, \cdots, \delta_{n,{N_{\rm RF} }} $ are test to minimize the objective. However, the number of combination is $ 2^{b{N_{\rm RF}}} $, which has prohibitively high computational complexity. For example, if $ b=4,{N_{\rm RF}}=32 $, the $ 2^{b{N_{\rm RF}}}  \approx 7.9\times10^{28}$! Thus, we need to investigate near-optimal search method to reduce complexity.
	
	In this case, we propose a high efficient individual search method, where each $  \delta_{n,i} $ is determined separately in each iteration. The specific procedures are summarized in \textbf{Algorithm 2}. We firstly initialize the $  \delta_{n,1}^{0}, \cdots, \delta_{n,{N_{\rm RF} }}^{0} $ by choosing the entry from the $ {\bf \Phi}_b $ and generate $ {\bf F}_{\rm RF}^0 $. In $ p $-th iteration, we find best $  \delta_{n,1}, \cdots, \delta_{n,{N_{\rm RF} }} $  one by one. In step 6, for $  \delta_{n,i} $,  we search through the $ {\bf \Phi}_b $ to find the optimal choice to satisfy the (\ref{P2_3}). This iterative process performs stop until the number of iterations reaches predetermined figure or  $ \delta_{n,i}^{p-1}= \delta_{n,i}^{p}$. Then the $ n $-th row of the designed $ {\hat{\bf F}}_{\rm RF} $ can be expressed as 
	\begin{equation}\label{FRF}
	\left[\hat{\bf{F}}_{\mathrm{RF}}\right]_{n,:}=\left[e^{j \hat{\delta}_{n,1}}, e^{j \hat{\delta}_{n,2}}, \ldots, e^{j \hat{\delta}_{n, N_{\rm{RF}}}}\right]
	\end{equation}
	
	After $ T_{max}  $ iteration, we can obtain the final practical codeword as
	\begin{equation}
	{\bf v}_p = {\bf F}_{\rm RF}^{T_{max}}{\bf f}_{\rm BB}^{T_{max}} 
	\end{equation}
	
	\begin{figure}[tbhp]
		\centering			
		\vspace*{0mm}\hspace*{-2mm}\includegraphics[width=3.9in]{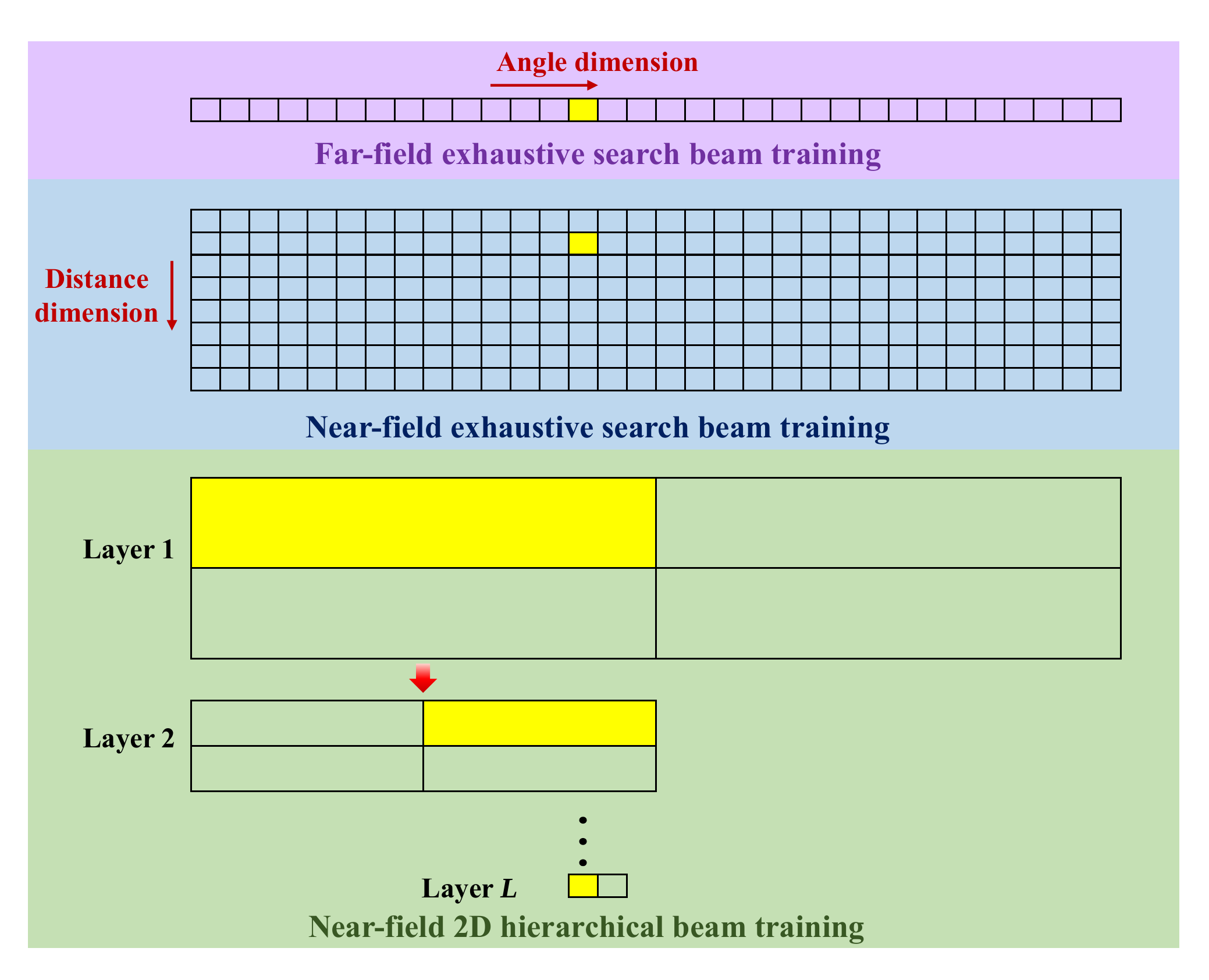}
		\centering
		\caption{Comparison between the far-field exhaustive search, near-field exhaustive search and the near-field 2D hierarchical beam training.}
		\vspace*{-2mm}
		\label{Scenario}
	\end{figure}
	
	\vspace{-0mm}
	
	\section{Proposed Near-Field 2D Hierarchical Beam Training}\label{S5}
	In this section, we first introduce the proposed near-field 2D hierarchical beam training scheme, where the angle and distance ranges are reduced gradually layer by layer in multi-resolution codebooks. Then, the analysis of the proposed beam training overhead is provided. 
	
	\vspace{-0mm}
	\subsection{Near-Field 2D Hierarchical Beam Training Scheme}
	In order to obtain the tradeoff between the near-field beam training overhead and the performance, one of the methods is to apply a hierarchical near-field codebook, which consists of multi-resolution codebooks. The sizes of codebooks are determined by the angle sample step and distance sample step of the codebook, i.e., $B_{\theta}  $ and $ B_{r} $ in (\ref{coverage}). 
	Specifically, as the increase of $B_{\theta}  $ and $ B_{r} $, the corresponding codeword has a lower resolution, and the size of the corresponding codebook becomes smaller. As mentioned before, we can generate near-field multi-resolution codebooks with different angle coverage and distance coverage based on the \textbf{Algorithm 1} and \textbf{Algorithm 2}. 
	
	Then, these multi-resolution codebooks are applied to conduct near-field 2D hierarchical beam training. Compared with far-field scenario, the near-field 2D hierarchical beam training need to reduce the search range of angle and distance at the same time as shown in Fig.~\ref{Scenario}.
	
	The specific near-field beam training procedure is summarized in \textbf{Algorithm 3}. First, as shown in Step2, for $l$-th codebook generation, we need to divide the angle coverage $ {\bf B}_{\mathbf{v}_{k},\theta}^l $ and distance coverage $ {\bf B}_{\mathbf{v}_k,r}^l $ based on angle samples step $ B_{\theta}^l  $ and distance samples step $ B_{r}^l  $ for each codeword $ \mathbf{v}_{k} $. Then, in Steps 3-4, the codewords design scheme based on \textbf{Algorithm 1} and \textbf{Algorithm 2} is applied to obtain the $ l $-th codebook $ {\bf{W}}_l $. Then, Steps 7-16 are operated to search the optimal codeword in multi-resolution codebooks layer by layer.
	\vspace{0mm}
	
	\begin{algorithm}[htbp]
		\caption{Near-field 2D hierarchical beam training}
		\textbf{Inputs}: $L$,$\left\lbrace {B_{\theta}^1, B_{\theta}^2,\cdots, B_{\theta}^L}\right\rbrace $, $\left\lbrace {B_{r}^1, B_{r}^2,\cdots, B_{r}^L}\right\rbrace $, $ {{y}_{opt}} =0$, $ {{s}_{opt}} =0$
		\\ // Generate $ L $ sub-codebooks 
		\\1. \textbf{for} $l = 1,2,\cdots, L$ \textbf{do}
		\\2. \hspace*{+3mm} generate the collection of $ {\bf B}_{\mathbf{v}_{l,k},\theta}^l $ and $ {\bf B}_{\mathbf{v}_{l,k},r}^l $ based on $ B_{\theta}^l  $ and $ B_{r}^l  $
		\\3. \hspace*{+3mm} generate $ \left| g_{\bf{v}}(\theta ,r)\right| $ for based on (\ref{coverage})
		\\4. \hspace*{+3mm} obtain the practical codewords in $ l $-th sub-codebook $ {\bf{W}}_l $ based on \textbf{Algorithm 1} and \textbf{Algorithm 2}.
		\\5. \textbf{end for}
		\\6. $ {\bf W} = {\bf W}_1 $
		\\ // Conduct beam training
		\\7. \textbf{for} $l = 1,2,\cdots, L$ \textbf{do}
		\\8. \hspace*{+3mm}\textbf{for} ${\bf v}_{l,k} $ in $ {\bf{W}} $  \textbf{do}   
		\\9. \hspace*{+6mm}$ {{y}_{k}^l} = {\bf{h}}^H{{\bf v}_{l,k}}s + {{n}}   $
		\\10. \hspace*{+6mm}\textbf{if} $ {{y}_{k}^l} > {{y}_{opt}}$ \textbf{then}
		\\11.\hspace*{+6mm}$ {{k}_{opt}} = k$
		\\12.\hspace*{+6mm}\textbf{end if}
		\\13.\hspace*{+3mm}\textbf{end for}
		\\14.\hspace*{+3mm}choose ${\bf v}_{l+1,k} $ in ${\bf W}_{l+1}$ satisfied $ {\bf B}_{\mathbf{v}_{l+1,k},\theta}^{l+1} \in {\bf B}_{\mathbf{v}_{l,k_{opt}},\theta}^l $ and ${\bf B}_{\mathbf{v}_{l+1,k},r}^{l+1} \in {\bf B}_{\mathbf{v}_{l,k_{opt},r}}^l $
		\\15.\hspace*{+3mm}the chosen codewords ${\bf v}_{l+1,k} $ compose the $ {\bf W}$ 
		\\16.\textbf{end for}
		\\\textbf{Output}: The feedback optimal codeword index $ k_{opt} $
		from the user.
		
	\end{algorithm}

	\subsection{Comparison of the Beam Training Overhead}
	Beam training overhead refers to the number of time slots used for beam training. Generally, the beam training overhead is determined by the spatial resolutions of an antenna array on the angle and distance, i.e., the number of sampled angle grids $ U $ and the number of sampled distance grids $ S $. It is worth pointing out that $  U $ is usually set as the same as the number of antennas on the array.
The training overhead of the exhaustive near-field beam training scheme is $ US $. Meanwhile, the training overhead of the time-delay based beam training is only related to the number of sampled distance grids $ S $. For the proposed 2D hierarchical beam training method, the beam training overhead can be represented as $ \mathcal{O}\left( log\left(U \right) +log\left(S \right) \right) $. It is obvious that, the training overhead of the proposed 2D hierarchical beam training is much less than that of the exhaustive near-field beam training. Since the number of sampled angle grids $ U $ is usually large than the number of sampled distances $ S $~\cite{Rainbow}, the training overhead of the proposed 2D hierarchical beam training is larger than that of the time-delay based beam training. However, the performance of the time-delay based beam training heavily depends on the extra hardware overhead and wideband condition, which will be further verified by simulation results in Section VI.
	
	\vspace*{-1mm}
	\section{Simulation Results}\label{S6}
	\vspace{0mm}
	For simulations, we assume that the number of BS antennas and RF chains are $N=512$ and $N_{\rm RF}=100$. The wavelength is set as $\lambda=0.005$ meters, corresponding to
	the $60$ GHz frequency. The quantified bits number of phase shifters is set as $b =  5 $. The path gain $\alpha$, angle $\theta$ and distance $r$ are generated as following:  $\alpha_l\sim{\cal CN}\left(0,1\right)$,  $\theta_l\sim{\cal U}\left(-1, 1\right)$,  and $r_l \sim{\cal U}\left(20, 100\right)$ meters. The SNR is defined as $1/{\sigma}^2$. 
	

\begin{figure}
	\centering
	\subfigure[Layer 1: Ideal beam pattern]{
		\begin{minipage}[t]{0.5\linewidth}
			\centering			
			\hspace*{-3.5mm}\includegraphics[width=2.5in]{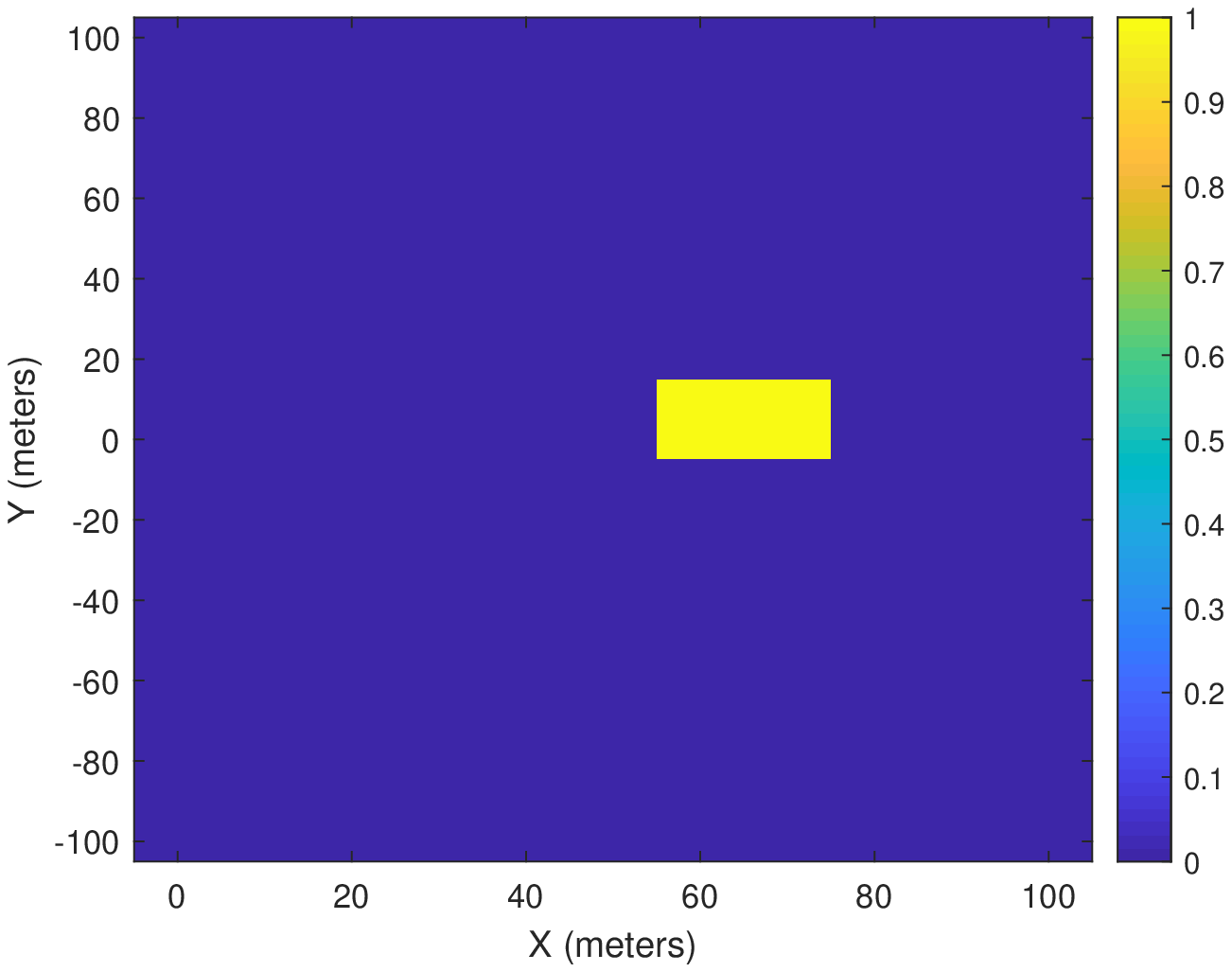}
		\end{minipage}
	}%
	\subfigure[Layer 1: Practical beam pattern]{
		\begin{minipage}[t]{0.5\linewidth}
			\centering
			\hspace*{-3.5mm}\includegraphics[width=2.5in]{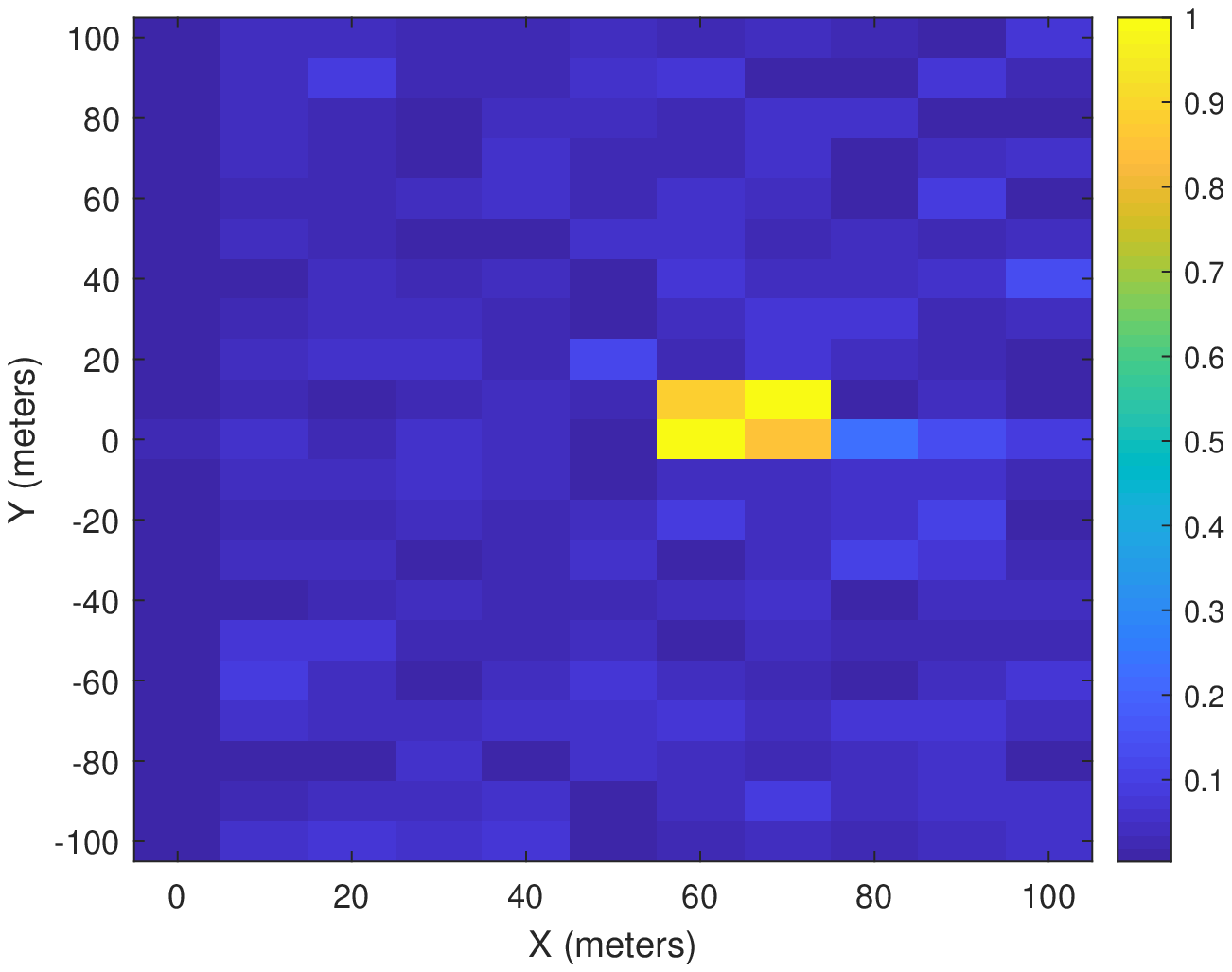}
		\end{minipage}
	}%
	
	\subfigure[Layer 2: Ideal beam pattern]{
		\begin{minipage}[t]{0.5\linewidth}
			\centering
			\hspace*{-3.5mm}\includegraphics[width=2.5in]{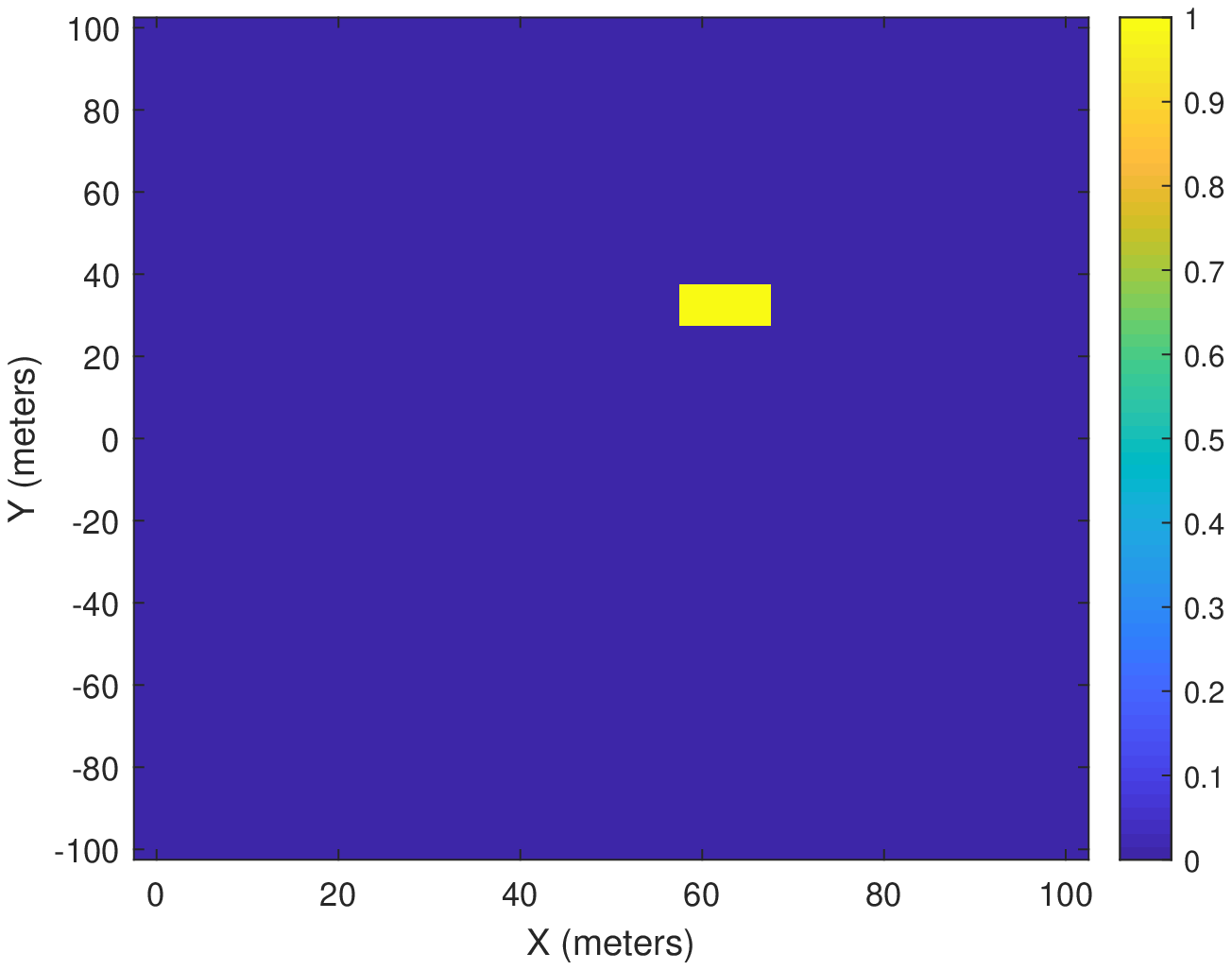}
		\end{minipage}
	}%
	\subfigure[Layer 2: Practical beam pattern]{
		\begin{minipage}[t]{0.5\linewidth}
			\centering
			\hspace*{-3.5mm}\includegraphics[width=2.5in]{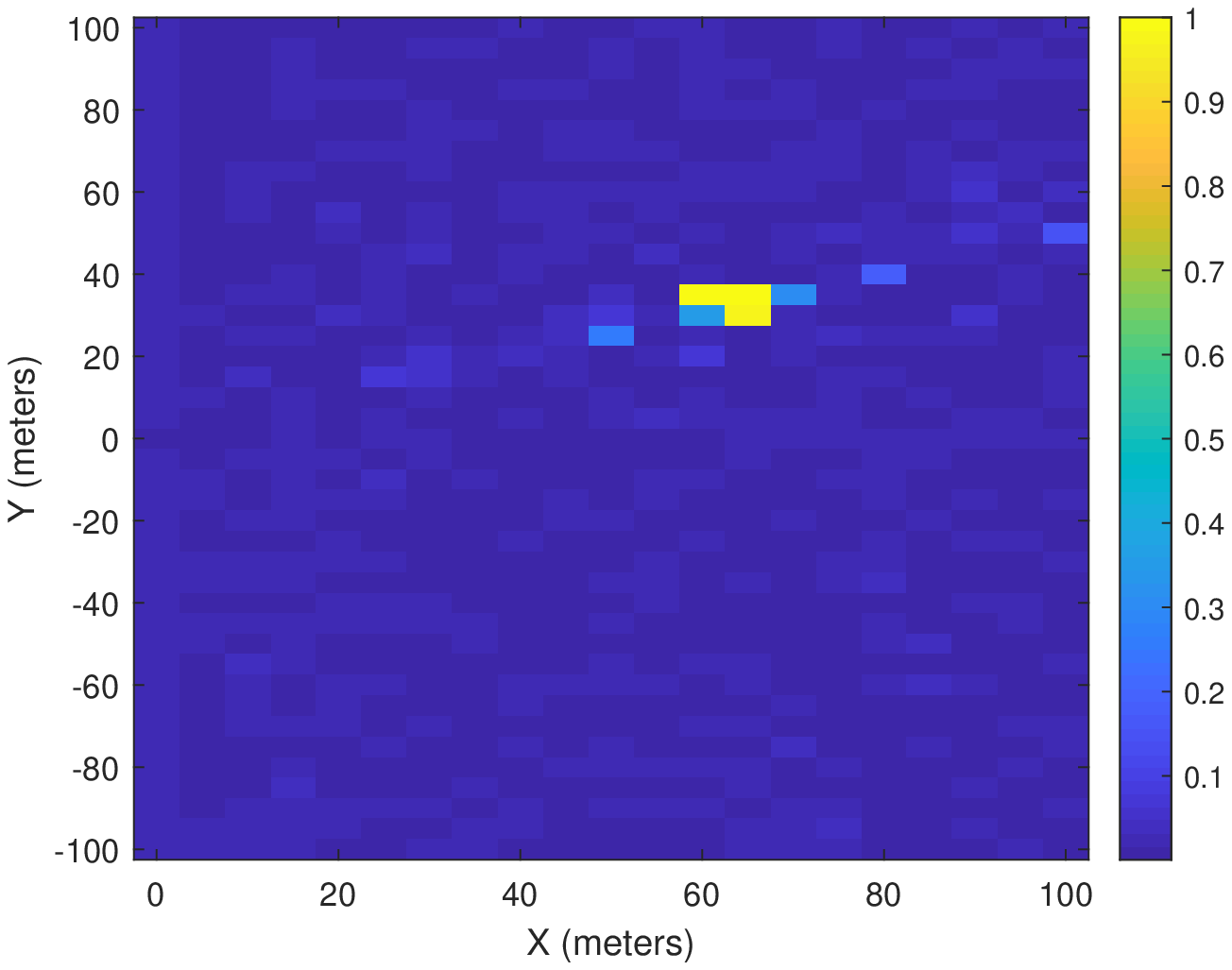}
		\end{minipage}
	}%
	\centering
	\vspace{-0mm}
	\caption{Comparison of the beam patterns of different layers of the hierarchical codebook.}\vspace{-0mm}
	\label{Sim1}
\end{figure}

	Fig. \ref{Sim1} shows the comparison of the ideal beam pattern and the normalized practical beam pattern obtained by conducting beamforming with the designed codeword. In these heat maps, the brighter the color, the greater the beamforming gain at this position. It is worth noting that, we utilize the rectangular coordinate system to present the beamforming gains of the locations in two-dimension space to show the beam pattern more clearly, where the coordinates of the X-axis and Y-axis satisfy $ x = r\cos(\theta) $, and $ y = r\cos(\theta) $.  Fig. \ref{Sim1} (a) presents an ideal beam pattern of the layer 1 codebook, where the beam should focus on the target location, i.e., $ x=[55,75],y=[-5,15] $. After we conduct beamforming with the designed practical codeword, we can obtain Fig. \ref{Sim1} (b), which presents the beamforming gains of different locations in space with the designed practical codeword. From Fig. \ref{Sim1} (b) we can see that the target location has the largest beamforming gain and other locations have much lower beamforming gains. Moreover, for the codeword in the layer 2 codebook, the designed practical codeword can also approach the ideal beam pattern Fig. \ref{Sim1} (c) and (d). Since the codeword in the layer 1 codebook should cover a larger range than that of layer 2 codebook, we can observe that the beamforming gain of non-target position in Fig. \ref{Sim1} (b) is also larger than that in Fig. \ref{Sim1} (d).

	\begin{table}[]	
		\renewcommand{\arraystretch}{1.3}
		\centering
		\caption{Comparisons of Beam Training Overhead}
		\vspace{-2mm}
		\label{Table1}
		\begin{tabular}{|c|l|l|}
			\hline
			Method                                        & Overhead & Value \\ \hline
			Far-field hierarchical scheme~\cite{far_hi}                & $ \sum_l^L U^{(l)}  $         & 40    \\ \hline
			Far-field exhaustive search scheme~\cite{OMP}  & U        & 512   \\ \hline
			Near-field exhaustive search scheme~\cite{Mingyao} & US       & 8192  \\ \hline
			Time-delay based near-field scheme~\cite{Rainbow}            & S        & 16    \\ \hline
			Proposed near-field 2D hierarchical scheme             & $ \sum_l^L U^{(l)} S^{(l)}  $       & 268   \\ \hline
		\end{tabular}
	\vspace{-2mm}
	\end{table}
	
	Table. \ref{Table1} presents the comparison of beam training overhead for different methods. We compare the proposed near-field 2D hierarchical beam training algorithm with the existing far-field hierarchical beam training scheme~\cite{far_hi}, far-field exhaustive search beam training scheme~\cite{OMP}, the near-field exhaustive search beam training scheme~\cite{Mingyao}, and time-delay based near-field beam training scheme~\cite{Rainbow}. We set the number of
	angle and distance grids as $ U = 512 $ and $ S = 16 $, respectively. The overhead of the far-field exhaustive search is set as the same as the number of sampled angle grids, i.e., $512$.
	The overhead of the near-field exhaustive search beam training scheme is set as $ 512 \times 16=8192 $. The overhead of time-delay based near-field beam training relates to the number of sampled distance grids, which is set as $16 $. For the far-field hierarchical beam training scheme, $U^{(l)} $ is the number of sampled angles in the $ l $-th layer, where $U^{(1)} = 4, U^{(2)} = 4, U^{(2)} = 32 $. Thus, the overhead of far-field hierarchical beam training is $ \sum_l^L U^{(l)} = 4+4+32 =40$. For the proposed near-field 2D hierarchical beam training algorithm, we use a three-layer hierarchical codebook. The size of the layer 1 codebook can be calculated as $ 64\times 4 = 256 $, where the numbers of sampled angle and distance grids are set as $ 64 $ and $ 4 $. For the layer 2 and layer 3 codebooks, we only need to search $ 8 $ and $4$ codewords. Thus the overhead of the proposed near-field 2D hierarchical beam training algorithm is $ 268 $, which is almost half of $ 512 $ and only 3.3 \% of $ 8192 $.
	
	
	\begin{figure}[tp]
		
		\vspace*{-2mm}
		\centering 
		\subfigure[]{
			
			\vspace*{0mm}\includegraphics[width=0.6\linewidth]{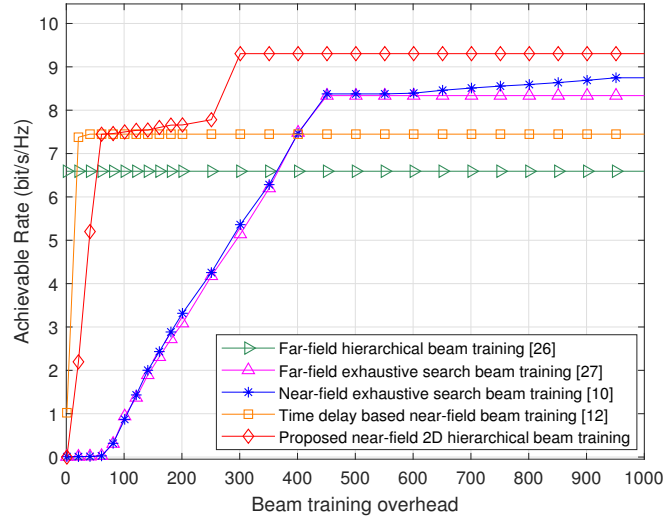}

		}\vspace*{-3mm}
		
		\subfigure[]{
			
			\vspace*{12mm}\includegraphics[width=0.6\linewidth]{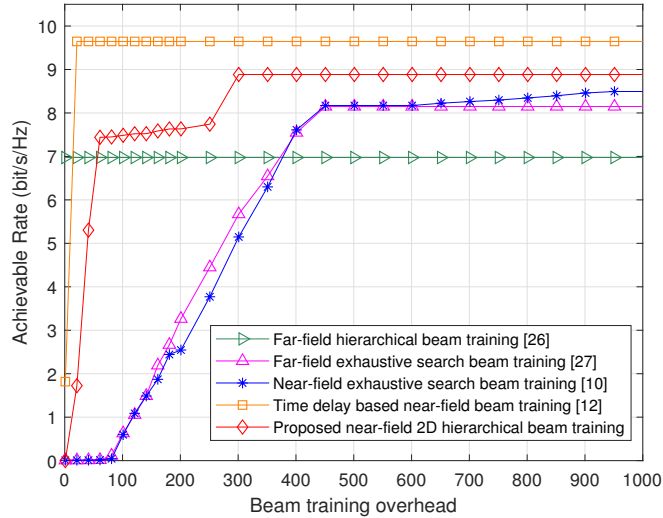}
			
		}
		\vspace*{-1mm}
		\caption{ \color{black}Achievable sum-rate performance comparison with respect to the beam training overhead under different bandwidths. (a) $ 100  $ MHz; (b) $ 500  $ MHz.}\vspace{-2mm}
		\label{FIG3}
	\end{figure}

	\begin{figure}[tp]
		
		\vspace*{-2mm}
		\centering 
		\subfigure[]{
			
			\vspace*{0mm}\includegraphics[width=0.6\linewidth]{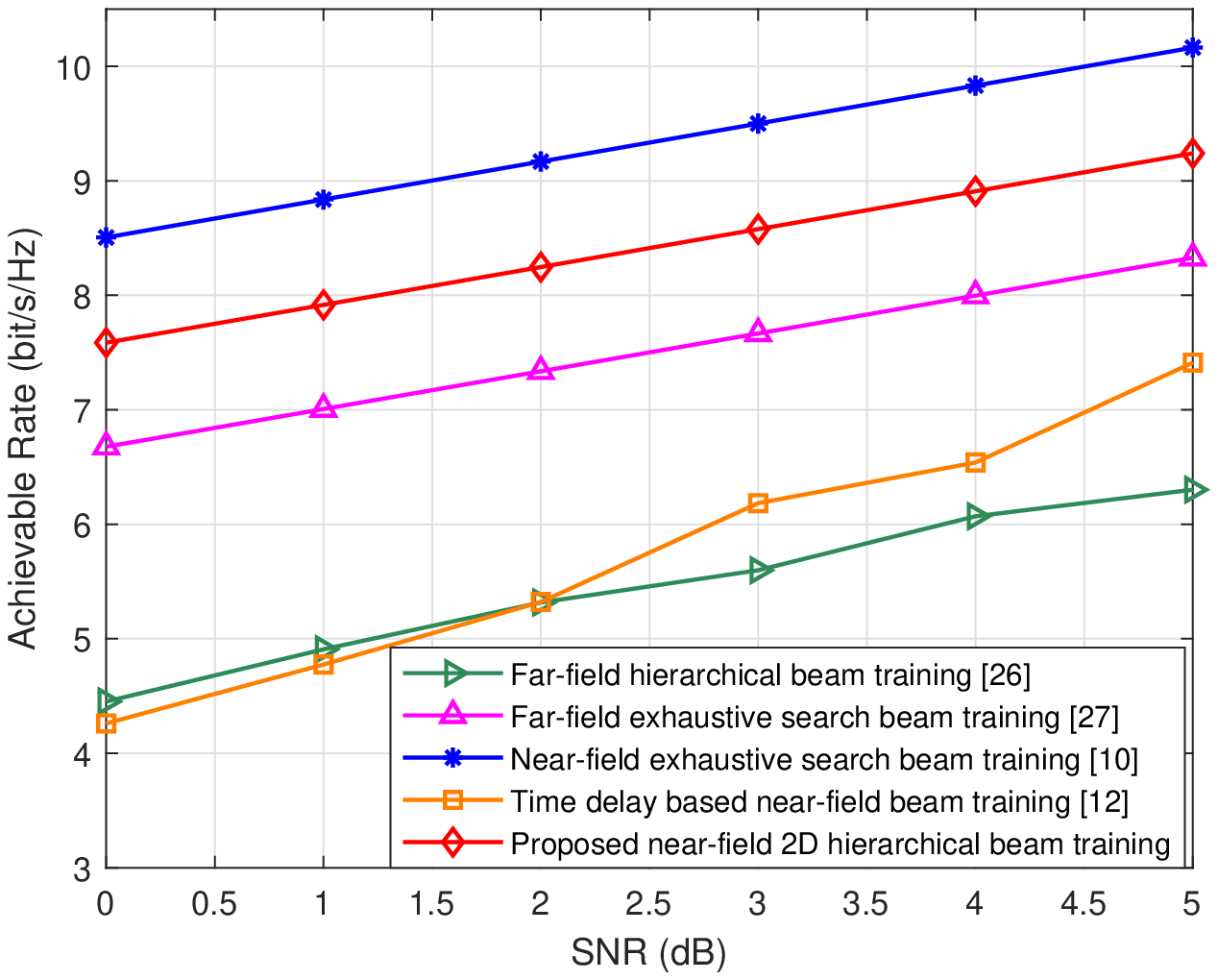}

		}\vspace*{-3mm}
		
		\subfigure[]{
			
			\vspace*{12mm}\includegraphics[width=0.6\linewidth]{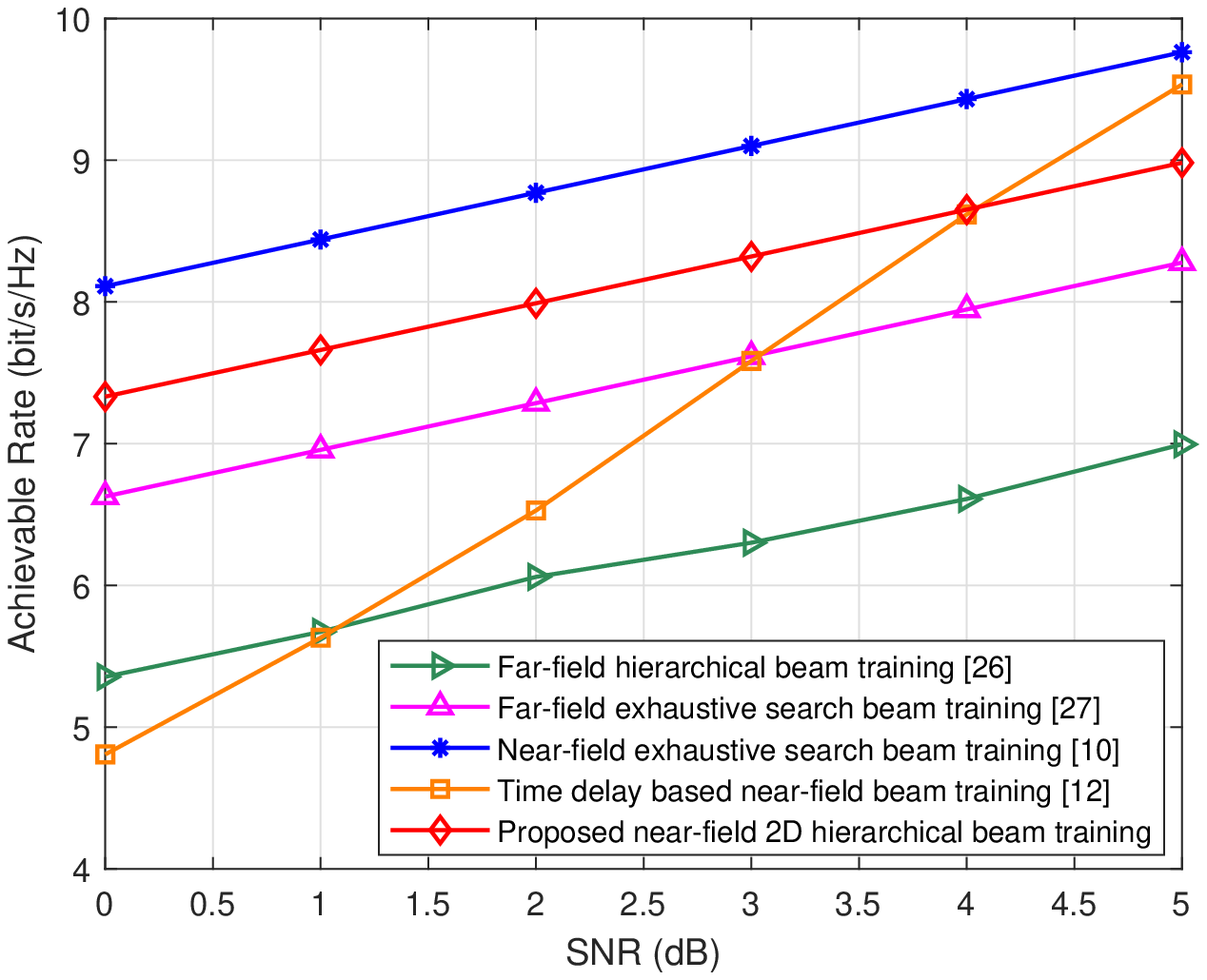}
			
		}
		\vspace*{-1mm}
		\caption{ \color{black}Achievable sum-rate performance comparison with respect to the SNR under different bandwidths. (a) $ 100  $ MHz; (b) $ 500  $ MHz.}\vspace{+0mm}
		\label{FIG4}
	\end{figure}
	
	Fig. \ref{FIG3} presents the performance of achievable rate comparisons against the beam training overhead under different bandwidths. The training overhead increases from 0 to 1000. In the beam training process, we utilize the optimal beamforming vector with the largest achievable rate searched in the current time slots to serve the user. 
	From Fig. \ref{FIG3} (a), where the bandwidth is $ 100  $ MHz, we can observe that the proposed near-field 2D hierarchical beam training can achieve the best performance of all schemes with relatively lower overhead. For example, the proposed scheme outperforms the far-field angle-domain codebook with only half of the beam training overhead. The reason is that the existing far-field codebook can only capture the angle information of the channel path. Moreover, the time-delay based scheme has worse performance than the proposed scheme in this narrow-band condition. The principal reason is that the ability of time-delay circuits to control the beam split will decrease by reducing the bandwidth. Meanwhile, Fig. \ref{FIG3} (b) illustrates the wideband situation, where the bandwidth is $ 500 $ MHz. It can be observed that the time-delay based beam training scheme has better performance than the proposed scheme. However, the proposed scheme has much lower hardware cost and is bandwidth-independent. Thus, we believe that the proposed scheme provides a tradeoff between the performance and overhead in near-field XL-MIMO beam training in a more general and cost-saving way.

	\begin{figure}[tp]
		
		\vspace*{-2mm}
		\centering 
		\subfigure[]{
			
			\vspace*{0mm}\includegraphics[width=0.6\linewidth]{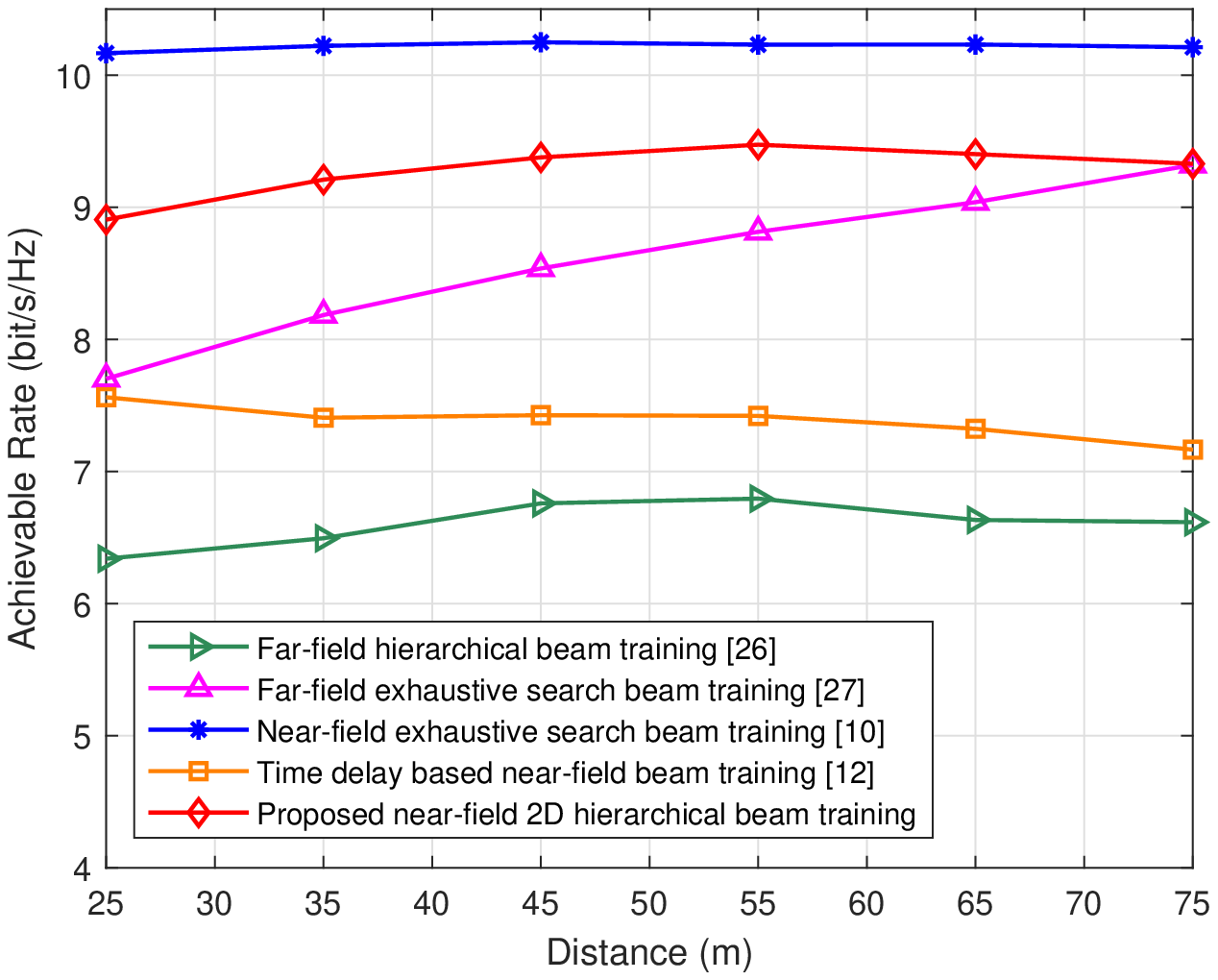}

		}\vspace*{-3mm}
		
		\subfigure[]{
			
			\vspace*{12mm}\includegraphics[width=0.6\linewidth]{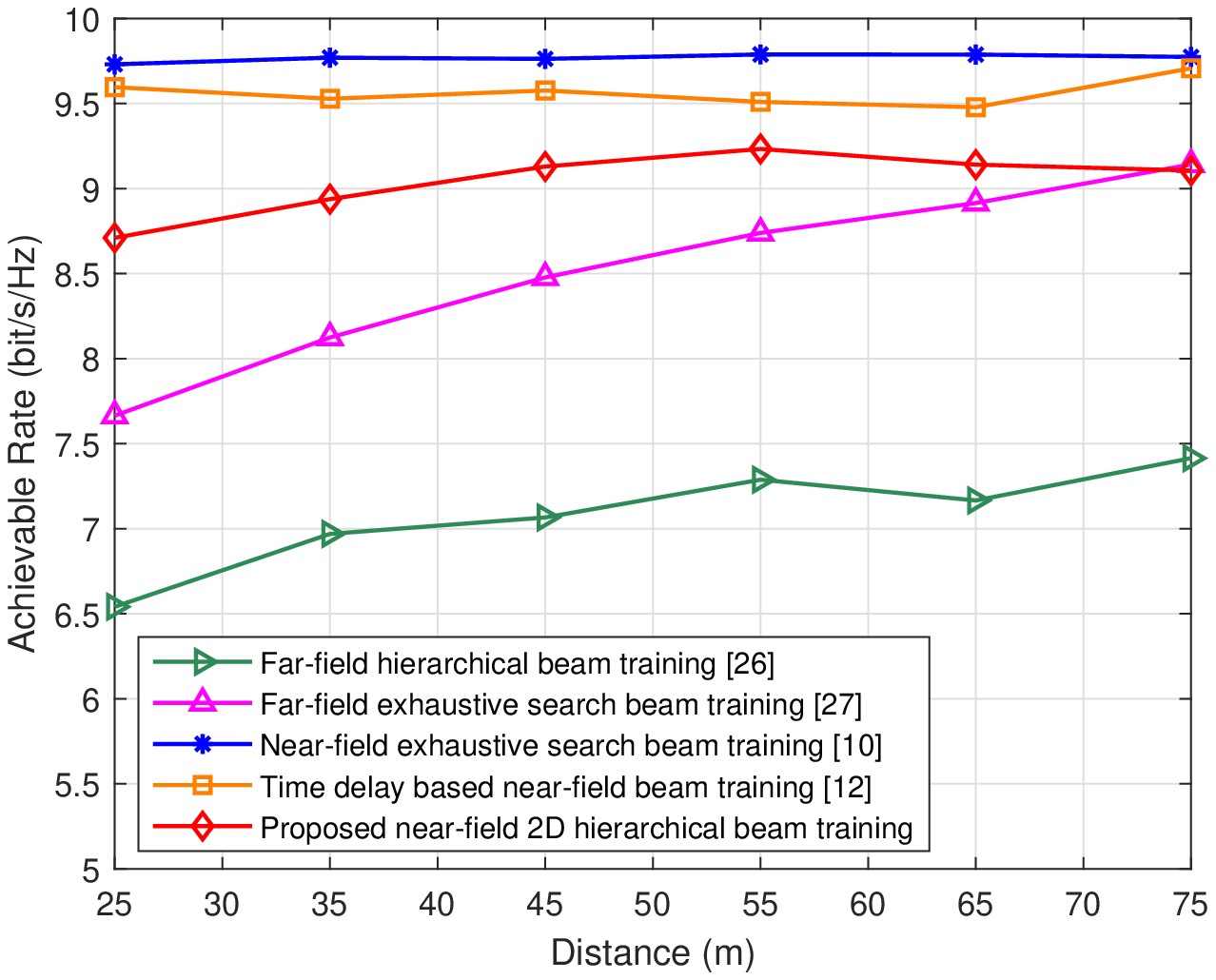}
			
		}
		\vspace*{-1mm}
		\caption{ \color{black}Achievable sum-rate performance comparison with respect to the distance overhead under different bandwidths. (a) $ 100  $ MHz; (b) $ 500  $ MHz.}\vspace{+0mm}
		\label{FIG5}
	\end{figure}
	
	Fig. \ref{FIG4} presents the performance of achievable rate comparisons against the SNR under different bandwidths, where SNR is from 0 dB to 5 dB. The simulation parameters are the same as those in Fig.~\ref{FIG3}. From Fig. \ref{FIG4} (a), i.e., narrow band condition, it is obvious that the proposed beam training scheme outperforms all existing far-field and near-field schemes. In specific, around 36.6\% improvement in achievable rate is accomplished by the proposed method compared to the time-delay based near-field beam training in SNR = 2 dB. 
	In addition, we can observe that the proposed method can also achieve better performance as long as SNR is smaller than 4 dB in the wideband situation. The reason why the near-field beam training scheme is vulnerable to noise is that the time-delay based near-field beam training scheme has to utilize beams with different frequencies to search different locations. the time-delay based near-field beam training scheme can not accumulate the power from all frequencies to combat noise as the near-field exhaustive beam training approach.

	Fig. \ref{FIG5} presents the performance of achievable rate comparisons against the distance under different bandwidths, where the distance is from 25 $\rm m $ to 75 $\rm m $ at SNR = 5 dB. From Fig. \ref{FIG5} (a), about 18.5\% performance improvement compared to the time-delay based near-field beam training at distance = 55 $ m $. Additionally, we can observe that the proposed method can also reach a 95.8\% achievable rate of the time-delay based near-field beam training at distance = 55 $ m $ in the wideband situation.
	
	\vspace*{0mm}
	\section{Conclusions}\label{S7}
	In this paper, we proposed a low-overhead near-field 2D hierarchical beam training by designing the near-field multi-resolution codebooks. Specifically, we first formulate the problem of designing near-field codeword and generating multi-resolution codebooks. It is worth pointing out that the proposed Gerchberg–Saxton (GS) based near-field codeword design algorithm can be utilized in designing codewords to realize arbitrary beam patterns. Then, a low-overhead near-field 2D hierarchical beam training scheme is proposed to realize the tradeoff between the training overhead and performance. Significantly, the proposed scheme can achieve sub-optimal performance without restriction to the hardware cost and wideband condition. 
	
	\vspace*{0mm}
	\begin{appendices}
		\section{Proof of \textbf{Lemma 1}}\label{AppendixA}
		The $ \| {\hat{\bf v}_{(s)}(u,w)\!-\!{\hat{\bf v}'_{(s)}(u,w) }} \|^{2}_{2} $ in (\ref{proof8}) can be further presented as 
		\begin{equation}\label{proof5}
		\begin{aligned}
		&\hspace{2mm}\| {\hat{\bf v}_{(s)}(u,w)\!-\!{\hat{\bf v}'_{(s)}(u,w) }} \|^{2}_{2} \\
		= &  \| {\hat{\bf v}_{(s)}(u,w) } \|_{2}^{2} + \|{\hat{\bf v}'_{(s)}(u,w) } \|_{2}^{2}\\
		&-2{\rm {Re}} \left( \left\langle {\hat{\bf v}_{(s)}(u,w) }, 
		{\hat{\bf v}_{(s+1)}(u,w)}  \right\rangle \right)\\
		= &\| {\hat{\bf v}_{(s)}(u,w) } \|_{2}^{2} + \|{\hat{\bf v}'_{(s)}(u,w) } \|_{2}^{2}\\
		&-2{\rm {Re}} \left( \|{\hat{\bf v}'_{(s)}(u,w) } \|_{2} \| {\hat{\bf v}_{(s)}(u,w) } \|_{2} \cos\phi \right)
		\end{aligned}
		\end{equation}
		where $ \phi $ is the angle between the $ \hat{\bf v}_{(s)}(u,w) $ 
		and $ {\hat{\bf v}}'_{(s)}(u,w)  $.
		Meanwhile, $ \| {\hat{\bf v}_{(s+1)}(u,w)\!-\!{\hat{\bf v}'_{(s)}(u,w) }} \|^{2}_{2} $ can be presented as 
		\begin{equation}\label{proof4}
		\begin{aligned}
		&\hspace{-22mm}\hspace{2mm} \| {\hat{\bf v}_{(s+1)}(u,w)\!-\!{\hat{\bf v}'_{(s)}(u,w) }} \|^{2}_{2} \\
		&\hspace{-25mm} =\| {\hat{\bf v}_{(s+1)}(u,w)} \|_{2}^{2} +\| {\hat{\bf v}'_{(s)}(u,w) } \|_{2}^{2} \\
		&\hspace{-22mm}-2 \| {\hat{\bf v}_{(s+1)}(u,w)} \|_{2} \| {\hat{\bf v}'_{(s)}(u,w)} \|_2\cos\tau.
		\end{aligned}
		\end{equation}
		where $ \tau $ is the angle between the $ {\hat{\bf v}}_{(s+1)}(u,w) $ 
		and $ {\hat{\bf v}}'_{(s)}(u,w) $.
		As shown in Step 8 of the \textbf{Algorithm 1}, where (\ref{v_nor}) presents the normalization of the $ {\hat{\bf v}'_{(s)}(u,w) } $, thus, $ \|{\hat{\bf v}_{(s)}(u,w) }\|_2 $ $ = $ $ \|{\hat{\bf v}_{(s+1)}(u,w) }\|_2 $ $ = $ 1, and the phase information of $ {\hat{\bf v}_{(s+1)}(u,w) } $ and $ {\hat{\bf v}'_{(s)}(u,w) } $ are the same, i.e., the angle between the $ {\hat{\bf v}}_{(s+1)}(u,w) $ 
		and $ {\hat{\bf v}}'_{(s)}(u,w) $ $ \tau $ is $ 0 $. In this case, (\ref{proof5})-(\ref{proof4}) is written as 
		\begin{equation}\label{proof6}
		\begin{aligned}
		&\hspace{2mm}\| {\hat{\bf v}_{(s)}(u,w)\!-\!{\hat{\bf v}'_{(s)}(u,w) }} \|^{2}_{2}-\| {\hat{\bf v}_{(s+1)}(u,w)\!-\!{\hat{\bf v}'_{(s)}(u,w) }} \|^{2}_{2} \\
		=& \|{\hat{\bf v}_{(s)}(u,w)}\|_2^2-\|{\hat{\bf v}_{(s+1)}(u,w)}\|_2^2\\
		&-2{\rm {Re}} \left( \|{\hat{\bf v}'_{(s)}(u,w) } \|_{2} \| {\hat{\bf v}_{(s)}(u,w) } \|_{2} \left( \cos\phi-\cos\tau\right)  \right) \\
		=&  2{\rm {Re}} \left( \|{\hat{\bf v}'_{(s)}(u,w) } \|_{2}\left( 1-\cos\phi\right)  \right)
		\end{aligned}
		\end{equation}
		Since $  \|{\hat{\bf v}'_{(s)}(u,w) } \|_{2}\left( 1-\cos\phi\right)  $ is always greater than zero, we can obtain that 
		\begin{equation}\label{proof7}
		\begin{aligned}
		\| {\hat{\bf v}_{(s)}(u,w)\!-\!{\hat{\bf v}'_{(s)}(u,w) }} \|^{2}_{2} > \| {\hat{\bf v}_{(s+1)}(u,w)\!-\!{\hat{\bf v}'_{(s)}(u,w) }} \|^{2}_{2}.
		\end{aligned}
		\end{equation}

	\end{appendices}
	
	\vspace*{3mm}
	\bibliography{IEEEabrv,LU}

\end{document}